\documentclass[preprint]{aastex61}
\submitjournal{ApJ}

\shorttitle{Faint Gamma-ray Pulsars}
\shortauthors{Smith et al.}
\begin{document}

\title{Searching a Thousand Radio Pulsars for Gamma-ray Emission}
%Author list created Wednesday 25 Sep 2018 \\
%%{\bf A few more eligible folks haven't signed up -- email Dave if wrong.}
\title{Searching a Thousand Radio Pulsars for Gamma-ray Emission}
\author[0000-0002-7833-0275]{D.~A.~Smith}  %  https://orcid.org/0000-0002-7833-0275
\email{smith@cenbg.in2p3.fr}
\affiliation{Centre d'\'Etudes Nucl\'eaires de Bordeaux Gradignan, IN2P3/CNRS, Universit\'e Bordeaux 1, BP120, F-33175 Gradignan Cedex, France}
\author[0000-0002-9032-7941]{P.~Bruel}  %  https://orcid.org/0000-0002-9032-7941
\email{Philippe.Bruel@llr.in2p3.fr}
\affiliation{Laboratoire Leprince-Ringuet, \'Ecole polytechnique, CNRS/IN2P3, F-91128 Palaiseau, France}
\author{I.~Cognard}
\affiliation{Laboratoire de Physique et Chimie de l'Environnement et de l'Espace -- Universit\'e d'Orl\'eans / CNRS, F-45071 Orl\'eans Cedex 02, France}
\affiliation{Station de radioastronomie de Nan\c{c}ay, Observatoire de Paris, CNRS/INSU, F-18330 Nan\c{c}ay, France}
\author{A.~D.~Cameron}
\affiliation{CSIRO Astronomy and Space Science, Australia Telescope National Facility, P.O. Box 76 Epping, NSW 1710, Australia}
\affiliation{Max-Planck-Institut f\"ur Radioastronomie, Auf dem H\"ugel 69, D-53121 Bonn, Germany}
\author[0000-0002-1873-3718]{F.~Camilo}
\affiliation{Square Kilometre Array South Africa, Pinelands, 7405, South Africa}
\author{S.~Dai}
\affiliation{CSIRO Astronomy and Space Science, Australia Telescope National Facility, P.O. Box 76 Epping, NSW 1710, Australia}
\author[0000-0002-9049-8716]{L.~Guillemot}   %   https://orcid.org/0000-0002-9049-8716 .
\affiliation{Laboratoire de Physique et Chimie de l'Environnement et de l'Espace -- Universit\'e d'Orl\'eans / CNRS, F-45071 Orl\'eans Cedex 02, France}
\affiliation{Station de radioastronomie de Nan\c{c}ay, Observatoire de Paris, CNRS/INSU, F-18330 Nan\c{c}ay, France}
\author{T.~J.~Johnson}
\affiliation{College of Science, George Mason University, Fairfax, VA 22030, resident at Naval Research Laboratory, Washington, DC 20375, USA}
\author[0000-0002-7122-4963]{S.~Johnston} %  https://orcid.org/0000-0002-7122-4963
\affiliation{CSIRO Astronomy and Space Science, Australia Telescope National Facility, P.O. Box 76 Epping, NSW 1710, Australia}
\author{M.~J.~Keith}
\affiliation{Jodrell Bank Centre for Astrophysics, School of Physics and Astronomy, The University of Manchester, M13 9PL, UK}
\author[0000-0002-0893-4073]{M.~Kerr}  %  http://orcid.org/0000-0002-0893-4073
\affiliation{Space Science Division, Naval Research Laboratory, Washington, DC 20375-5352, USA}
\author[0000-0002-4175-2271]{M.~Kramer}  % 0000-0002-4175-2271
\affiliation{Max-Planck-Institut f\"ur Radioastronomie, Auf dem H\"ugel 69, D-53121 Bonn, Germany}
\affiliation{Jodrell Bank Centre for Astrophysics, School of Physics and Astronomy, The University of Manchester, M13 9PL, UK}
\author[0000-0002-4799-1281]{A.~G.~Lyne}  %  0000-0002-4799-1281
\affiliation{Jodrell Bank Centre for Astrophysics, School of Physics and Astronomy, The University of Manchester, M13 9PL, UK}
\author[0000-0001-9445-5732]{R.~N.~Manchester}  % 0000-0001-9445-5732
\affiliation{CSIRO Astronomy and Space Science, Australia Telescope National Facility, P.O. Box 76 Epping, NSW 1710, Australia}
\author[0000-0002-7285-6348]{R.~Shannon}  %  0000-0002-7285-6348
\affiliation{ARC Centre of Excellence for Gravitational Wave Discovery, Centre for Astrophysics and Supercomputing, Swinburne University of Technology, Hawthorn, VIC 3122, Australia}
\author[0000-0002-8950-7873]{C.~Sobey} % https://orcid.org/0000-0002-8950-7873 
\affiliation{CSIRO Astronomy and Space Science, PO Box 1130, Bentley, WA 6102, Australia}
\affiliation{International Centre for Radio Astronomy Research - Curtin University, GPO Box U1987, Perth, WA 6845, Australia}
\author[0000-0001-9242-7041]{B.~W.~Stappers}  %   https://orcid.org/0000-0001-9242-7041  
\affiliation{Jodrell Bank Centre for Astrophysics, School of Physics and Astronomy, The University of Manchester, M13 9PL, UK}
\author[0000-0003-2122-4540]{P.~Weltevrede}  %   My ORCID number is: 0000-0003-2122-4540
\affiliation{Jodrell Bank Centre for Astrophysics, School of Physics and Astronomy, The University of Manchester, M13 9PL, UK}
\begin{abstract}
Identifying as many gamma-ray pulsars as possible in the {\em Fermi} Large Area Telescope (LAT) data helps 
test pulsar emission models by comparing predicted and observed properties for a large, varied sample  %  of observed pulsars
with as little selection bias as possible.
It also improves extrapolations from the observed population to estimate the contribution of unresolved pulsars to the diffuse gamma-ray emission.
We use a recently developed method to determine the probability that a given gamma-ray photon comes from a known position in the sky, 
convolving the photon's energy with the LAT's energy-dependent point-spread-function (PSF),
without the need for an accurate spatial and spectral model of the gamma-ray sky around the pulsar.
The method is simple and fast and, importantly, provides probabilities, or {\em weights}, for gamma rays from pulsars too faint for phase-integrated detection.
We applied the method to over a thousand pulsars for which we obtained rotation ephemerides from radio observations,
and discovered gamma-ray pulsations from 16 pulsars, 12 young and 4 recycled. 
PSR J2208+4056 has spindown power $\dot E = 8\times 10^{32}$ erg s$^{-1}$, about three times lower
than the previous observed gamma-ray emission ``deathline''. PSRs J2208+4056 and J1816$-$0755 have radio interpulses,
constraining their geometry and perhaps enhancing their gamma-ray luminosity.
We discuss whether the deathline is an artifact of selection bias due to the pulsar distance.
\end{abstract}

\keywords{gamma rays: observations -- pulsars: individual (J0636+5129, J1731$-$4744, J1816$-$0755, J2208+4056)}   

% 3 Dec -- AAS editors say "six max".
%\keywords{catalogs -- gamma rays: observations -- pulsars: general -- pulsars: individual (J0636+5129, J0729$-$1836, J1125$-$6014, J1327$-$0755, J1731$-$4744, 
%J1740+1000, J1757$-$2421, J1816$-$0755, J1841$-$0524, J1853$-$0004, J1913+1011, J1925+1720, J1928+1746, J1932+2220, J1946+3417, J2208+4056)-- stars: neutron }   

\section{Introduction} \label{sec:intro}
The lowest photon fluxes among the 117 gamma-ray pulsars reported in the Second {\em Fermi} Large Area Telescope (LAT) Catalog of Gamma-ray
Pulsars \citep[][hereafter ``2PC'']{2PC} were of order $F_{100}^\mathrm{min} \simeq 10^{-9}$ photons cm$^{-2}$ s$^{-1}$, integrated above 100 MeV. 
The LAT's effective area averaged over the energy range where pulsars are bright is $\sim 5000$ cm$^2$ \citep{LATinstrument}. 
{\em Fermi}'s continuous all-sky survey covers every point on the sky with a roughly $1 \over 6$ duty cycle. 
Combining, we find that $F_{100}^\mathrm{min}$ yields a modest average of 2 photons per month recorded by the LAT,
with $>100$\% Poisson fluctuations in the monthly counting rate. 
%>>> 1.e-9*8000.*30.*86400./6.   ==>  3.4560

Detecting lower-flux pulsars gives access to a larger space volume, containing a rich variety of pulsars.
Generally, luminosity scales with spindown power, $L_\gamma \propto \sqrt{\dot E}$ (see for example 2PC Fig. 9). 
However the dispersion around this trend spans over two orders of magnitude, presumably due to how the size, shape and
density of the emitting region depend on the neutron star's spin and magnetic field configuration.
Luminosity may thus be intrinsically low for pulsars with rare combinations of properties, 
or may appear low because only a shoulder of the gamma-ray beam sweeps across the Earth \citep{subluminous}.
``Variety'' therefore includes pulsars probing the $\dot E = 4 \pi^2 I \dot P P^{-3}$ deathline for gamma-ray emission.
$P$ is the period, $\dot P = {dP \over dt}$, and we set the moment of inertia to $I = 10^{45}$ g cm$^2$.

% Above re-write of below paragraph sparked by Regina Caputo comment, 22 October 2018.
%
%Detecting low-flux pulsars gives access to a large space volume, containing a rich variety of pulsars.
%Luminosity scales with spindown power, $L_\gamma \propto \sqrt{\dot E}$ (see for example 2PC Fig. 9), so  ``variety'' includes
%pulsars probing the $\dot E = 4 \pi^2 I \dot P P^{-3}$ deathline for gamma-ray emission.
%Luminosity may be intrinsically low for rare pulsars, 
%or the flux may be low because a shoulder of the gamma-ray beam sweeps across the Earth \citep{subluminous}.
%$P$ is the period, $\dot P = {dP \over dt}$, and we set the moment of inertia to $I = 10^{45}$ g cm$^2$.

Pulsars may be ``faint'', i.e. near the LAT's detection threshold, even for high count rates: 
background is intense for low Galactic latitudes and in confused regions.
Broad pulse widths also degrade sensitivity \citep{SixWeak}.
Detecting and characterizing faint pulsars helps ensure that emission models are confronted with an accurate snapshot of the true pulsar population. 
It also helps determine the pulsar luminosity function needed to quantify the contribution of unresolved pulsars to the diffuse emission, 
for example in the inner galaxy \citep{LATpsrGeVexcess}.
%Twenty-one gamma-ray pulsars found by phase-folding with radio rotation ephemerides after 2PC were presented by \citet{LaffonNewPSRs} and by \citet{DasAmsterdam}. 

We have been phase-folding LAT gamma-ray photons using hundreds of rotation ephemerides provided by the
astronomers of the Pulsar Timing Consortium \citep{TimingForFermi} since {\em Fermi} began routine observations on 2008 August 4 (MJD 54682).
The Consortium provided ephemerides for $\sim 95$\% of known pulsars with spindown power $\dot E > 10^{34}$ erg s$^{-1}$, and about half of those known with lower $\dot E$. 
This led to the discovery of gamma pulsations from roughly half of LAT's over 240 gamma-ray pulsars\footnote{\url{https://confluence.slac.stanford.edu/display/GLAMCOG/Public+List+of+LAT-Detected+Gamma-Ray+Pulsars}}. 
The rest came from searches at the positions of pulsar-like unidentified LAT sources: 
a quarter of the total are radio-quiet pulsars found in blind gamma-ray periodicity searches \citep[see][and references therein]{ClarkEatHomeI}, 
and nearly 90 are millisecond pulsars (MSPs) discovered via deep radio exposures \citep[see e.g.][]{LOFAR_LAT_MSP}.
But flagging a source as pulsar-like means it is bright enough in gamma rays to find its spectral shape and/or variability.
The \textit{faint} gamma-ray pulsars were all found by phase-folding with an ephemeris.

{\em Weighting} \citep{KerrWeighted} is a key tool for these searches:
the spatial and spectral map of the gamma rays around the pulsar's position, combined with the LAT's energy-dependent Point Spread Function (PSF), translates into the probability
of a given photon being signal or background (see Section \ref{WeightSection}). 
What is new in the present work is that we applied the simplified weighting method of \citet{SearchPulsation}
to discover even fainter objects, for which a phase-integrated source is often not even detectable.

In Section \ref{WeightSection} we summarize the \citet{SearchPulsation} photon weighting method. 
% , particularly well-suited for sources near the sensitivity threshold, for which the spectral determination needed for previous weighting methods is difficult or impossible.
The new photon weighting scheme is fast, making it useful for brighter sources as well.
In Section \ref{FoldingSection} we use the method to phase-fold 1269 pulsars, over 240 of which were known to pulse in gamma rays as this work was being completed.
The large sample highlights the stability and robustness of our search method, 
in particular that the significance ``noise floor'' is safely below $4\sigma$, lower than the $5\sigma$ threshold used in previous work.
Gamma-ray pulsations from 16 previously-known energetic radio pulsars are presented in Section \ref{FoldingSection}.
Folding the large sample also pushes the apparent gamma-ray emission ``deathline'' for pulsars to
spindown power $\dot E \lesssim 8\times 10^{32}$ erg s$^{-1}$ (Section \ref{DeathSection}). 
We summarize in Section \ref{ConclusionsSection}.

\begin{figure}[ht!]
\centering
\includegraphics[width=0.75\textwidth, angle=270]{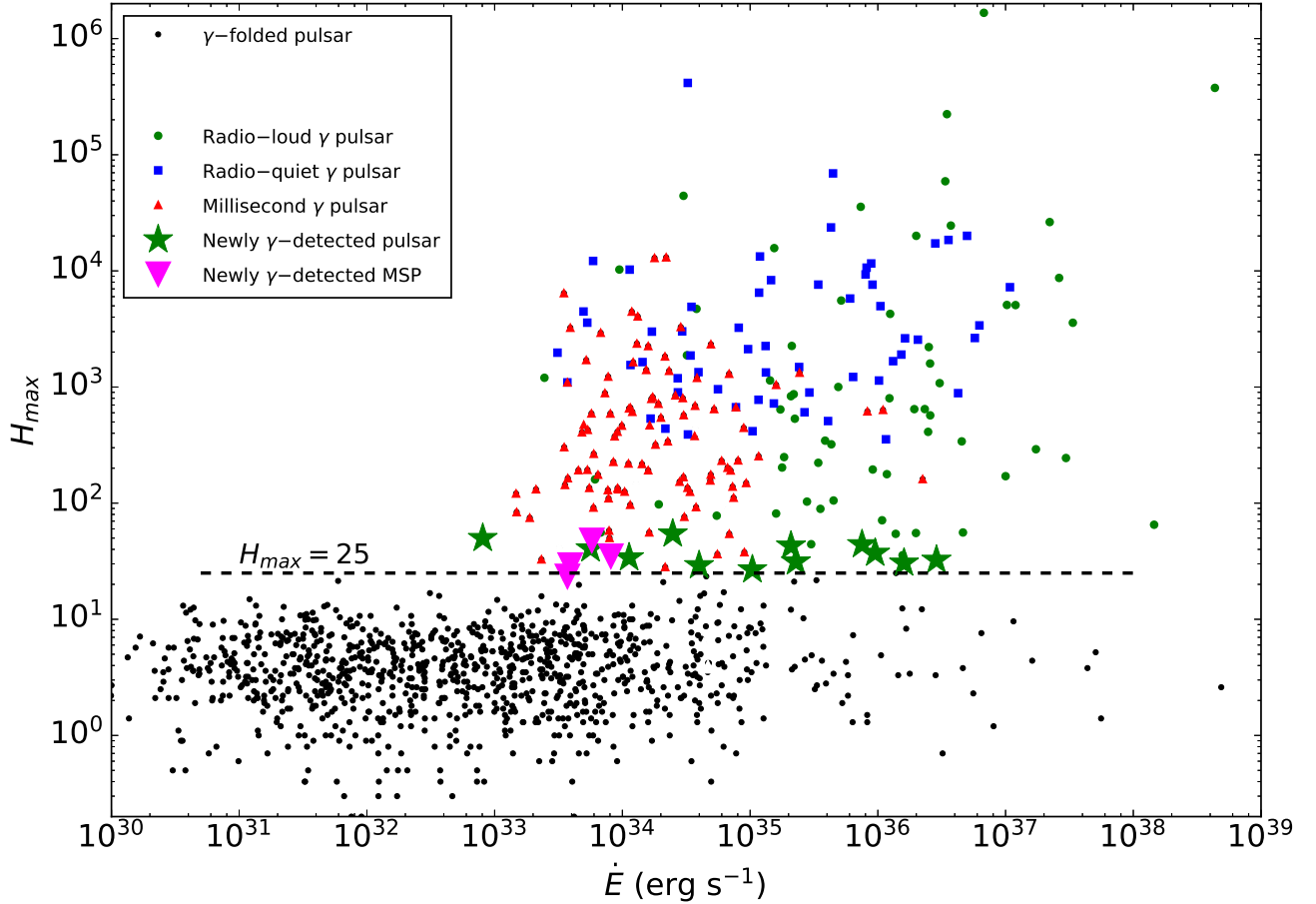}
\caption{$H_\mathrm{max}$ (see Section \ref{FoldingSection}) versus the spindown power $\dot E$ for 1269 pulsars, for $9.6$ years of LAT data.
Colored symbols indicate those known to pulse in LAT gamma rays.
Small black dots show pulsars that were gamma-ray phase-folded, for which pulsations are not detected.
The Shklovskii $\dot E$ correction for proper motion has not been applied here.
The 16 gamma-ray pulsars discovered in the course of this work are highlighted.
\label{H36_EDOT}}
\end{figure}

\section{Weighted Photon Folding, revisited}  \label{WeightSection}
A gamma ray crosses the vacuum of space for thousands of years without incident, to enter perhaps into the LAT.
Recoil by the atoms therein ensures conservation of momentum, allowing pair production, $\gamma \rightarrow e^+e^-$.
The electron and positron ionize the silicon strips along their path in the LAT's tracker, but undergo multiple Coulomb scattering,
mainly in the tungsten foils. 
Coulomb scattering causes the fundamental limit on the angular resolution of the LAT at low energy,
and determines its PSF. The spacing between the silicon strips limits the minimum PSF above 10 GeV.
\citet{das_djt_psr_detection} provide further discussion.

At the end of the complex chain of LAT event reconstruction and selection, 
the LAT's energy-dependent PSF is modeled as a Moffat distribution (see Eq. \ref{moffat}, below) in $\Delta \theta$,
the angular distance between a target direction (here, the pulsar's radio timing position) and a photon's sky direction $\vec \theta$.
The distribution's width in degrees (68\% containment) is
\begin{equation}
\sigma_\mathrm{psf}(E_\gamma) = \sqrt{p_0^2(E_\gamma/100)^{-2 p_1} + p_2^2},
\end{equation}
where $E_\gamma$ is the gamma-ray photon energy in MeV. 
For the P305 Pass 8 data\footnote{\url{https://fermi.gsfc.nasa.gov/ssc/data/} \label{fsscfootnote}} used in this work \citep{Pass8},
the parameter values are $p_0 = 5.445, p_1 = 0.848$, and $p_2 = 0.084$.

To search for a pulsed gamma-ray signal, we select gamma-ray photons within $5^{\circ}$ of the pulsar position,
approaching $\sigma_\mathrm{psf}$ for our $E_\gamma > 50$ MeV data sample.
Their arrival times in the LAT are converted to a rotational phase of the neutron star using
the parameters in the rotation ephemeris for that pulsar
and the {\tt fermi} plug-in \citep{Ray2011} to {\textsc Tempo2} \citep{Hobbs2006}.
We evaluate pulsation significance for the photon list with the H-test \citep{DeJager2010}.
We stack (fold) the phases in histograms to examine and analyze the profile.

In the presence of background photons from nearby sources and diffuse emission, 
the signal-to-noise ratio can be optimized using different methods.
The most effective to date is {\em weighting} \citep{KerrWeighted}.
The weight $w(E_\gamma, \vec\theta)$ is the probability that a photon with energy $E_\gamma$ and apparent sky direction $\vec \theta$
comes from the pulsar rather than background.
Kerr calculates the weights as $w = R_p/(R_b + R_p)$,
where $R_p$ is the rate of gamma rays from the pulsar,
and $R_b$ is the rate for gamma rays from all the background components.
The {\em Fermi} Science Tool {\tt gtsrcprob}\textsuperscript{\ref{fsscfootnote}} implements the method.
It takes as input a {\em ``source model''} describing the spectra and positions of all nearby gamma-ray sources,
created using the {\tt gtlike} tool, as well as $\sigma_\mathrm{psf}(E_\gamma)$ and $\vec \theta$, to calculate the expected rates and thus $w$.
Kerr modified the H-test to include weights, summarized succintly in Section 3 of 2PC. 
Weighting for the 117 pulsars in 2PC, and for nearly all LAT team pulsar studies since 2011, was done with {\tt gtsrcprob}.

However, the {\tt gtsrcprob} tool has two significant drawbacks.
First, {\tt gtlike} requires high detection significance to allow {\tt gtsrcprob} weights to be meaningful.
The analysis can be complex for faint sources in highly confused regions,
and may fail for faint sources where pulsations are nevertheless clearly detectable. 
Second, {\tt gtlike}, and the tools {\tt gtdiffrsp} and {\tt gtexposure} that its use requires, 
impose much larger data files and longer computation times than needed for the simple weighting method.

\subsection{Simplified Weighting}
\label{weightsubsect}
\begin{figure}[ht!]
\centering
\includegraphics[width=0.9\textwidth]{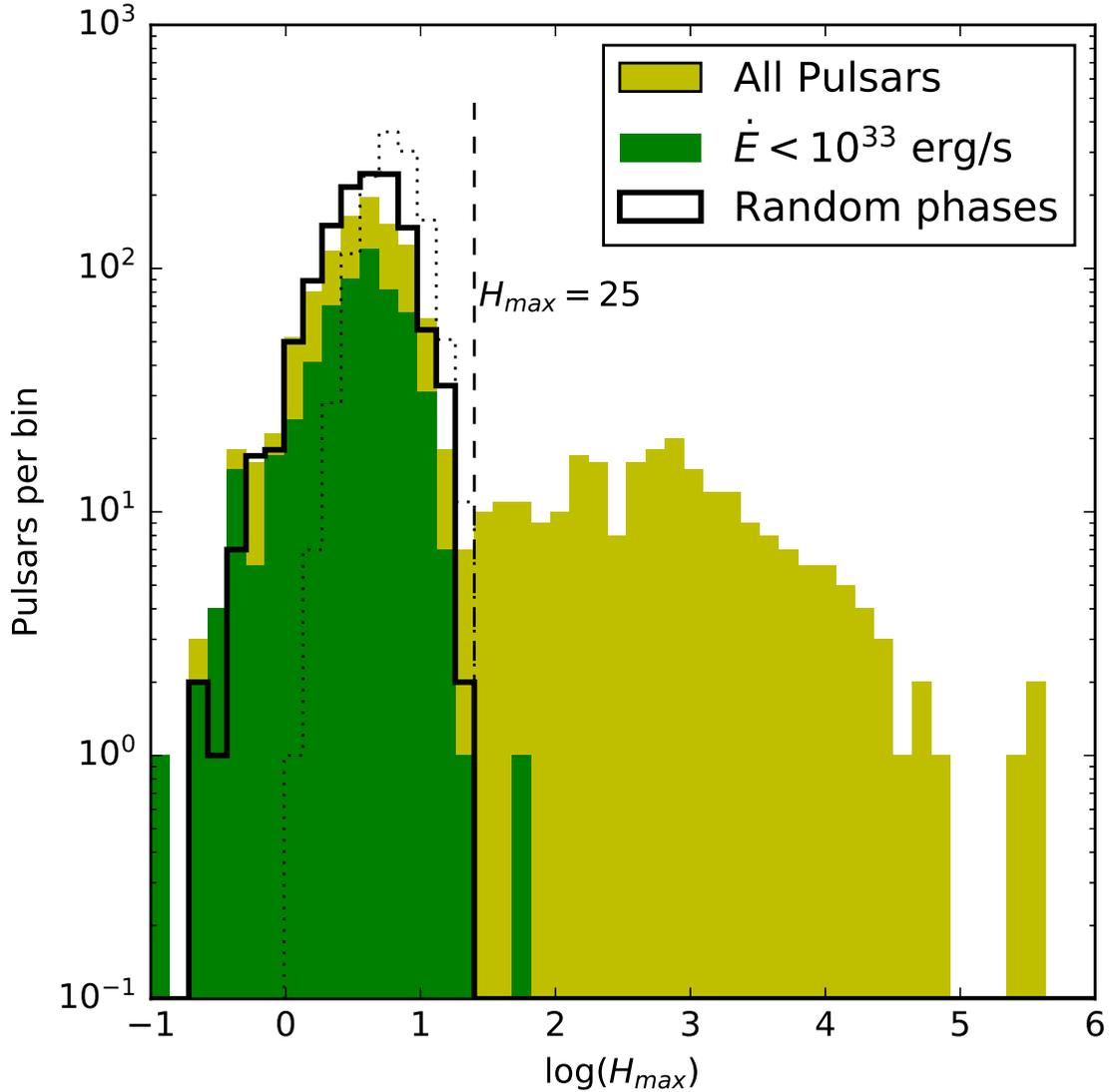}
\caption{Histograms of the $H_\mathrm{max}$ values shown in Figure \ref{H36_EDOT},
and of the maximum H-test values obtained for 3 (solid line) and 8 (dotted line) sets of random phases generated for each pulsar's photon list (see Section \ref{FoldingSection}). 
\label{HistoHs}}
\end{figure}

\citet{SearchPulsation} provides a simple formula that approximates the 
weights remarkably well,\footnote{A python script to calculate the weights is at \url{https://fermi.gsfc.nasa.gov/ssc/data/analysis/user},
and also at \url{https://github.com/nanograv/PINT/blob/master/pint/fermi_toas.py} .}
\begin{equation} \label{weightEq}
w(E_\gamma, \Delta \theta)  = g(E_\gamma, \Delta \theta) \exp ( {-2\log^2(E_\gamma / E_\mathrm{ref})}),  
\end{equation}
with one free parameter for the energy scale, $\mu_w = \log_{10} (E_\mathrm{ref}/ 1\,\mathrm{MeV})$, discussed below.
The geometrical factor $g$ uses a Moffat distribution \citep{Moffat}\footnote{Mis-attributed to \citet{KingFunction}
in \citet{xmmKingFunction}, corrected in the LAT analysis documentation.}
to describe the angular distribution of the gamma-ray photons emanating from a point source,
\begin{equation}
g(E_\gamma, \Delta \theta) = \left(1+ {9\Delta \theta^2 \over 4\sigma_\mathrm{psf}^2(E_\gamma)} \right)^{-2}.
\label{moffat}
\end{equation}
The Moffat distribution resembles a Gaussian but with broader tails.

The exponential factor was obtained by studying weight distributions computed under the simple assumptions that the pulsar rate is much lower than the background rate, 
and that the background is uniform. \citet{SearchPulsation} explores the validity of these assumptions. Pulsar spectra are of the form
\begin{equation}\label{expcutoff}
\frac{{\rm d} N}{{\rm d} E_\gamma} = K \Big(\frac{E_\gamma}{1\,\mathrm{GeV}}\Big)^{-\Gamma} \exp \left(- \frac{E_\gamma}{E_{\rm cut}} \right)
\end{equation}
that is, hard spectra (typically $0.5 < \Gamma < 2.0$) with sharp cut-offs ($0.6 < E_{\rm cut} < 6$ GeV, see 2PC). K determines the flux. 
In the energy range germane to pulsars, the spectrum of the diffuse background is a steeper power law. 
In these conditions, the pulsar weights versus $\log E_\gamma$ are well-approximated by Gaussians with width $0.5$, relatively independent of the pulsar's spectrum. 
%Neighboring sources (mainly other pulsars, supernova remnants, pulsar wind nebula, and blazars) further contribute to the background
%local to a given pulsar and determine the background's overall spectral shape.
%
%The simulated distributions of pulsar weights versus $\log E_\gamma$ for a pulsar immersed in background are well-approximated by gaussians with 
%width $\log \sigma_w = 0.5$, for energies expressed in MeV, relatively independent of the pulsar's spectrum or environment. 
What changes from one pulsar to another is the reference energy $E_\mathrm{ref} = 10^{\mu_w}$ for which the pulsar's weight distribution peaks.
For a faint pulsar with a soft spectrum far off the Galactic plane (hence, with low background), $E_\mathrm{ref}$ can be below 1 GeV ($\mu_w = 3$).
A pulsar with a hard spectrum in a highly confused region can have a distribution peaking above 10 GeV ($\mu_w = 4$).

\subsection{Choosing $\mu_w$} \label{choosemu}
For a detectable gamma-ray pulsar, H-test versus $\mu_w$ follows a bell-like curve. 
To find $E_\mathrm{ref}$ that maximizes pulsation significance, \citet{SearchPulsation} describes a simple procedure to
find the curve's peak with the fewest trials.
% and yet ensure a small, well-defined number of search trials, 
%one can calculate the H-test statistic for, say, five $\mu_w$ values, fit a curve, set $\mu_w$ to the curve's peak value, and correct the significance for the number of trials.
%The problem remains that for faint pulsars, or for pulsars for which the rotation ephemeris turns out not to phase-fold gamma-rays properly
%over its nominal validity interval, this procedure can yield a wrong $\log E_\mathrm{ref}$ value and a missed detection.
Here, we proceeded differently. 
After phase-folding many gamma-ray pulsars, we saw that for most, $\mu_w = 3.6$ yields a significant signal. 
%\footnote{PSR J1838-0655 that I looked at with Frank Marshall and Gabriele
%Brambilla likes $\log E_\mathrm{ref} = 1.9$ (!) and requires photons below 100 MeV. If that paper exists when this one is nearing completion, we can cite it.}
The values of $\mu_w = 3.2$ and $4.0$ work for the rest.
After detection, the above fitting procedure tunes $\mu_w$ to optimize the H-test significance.
Often, to confirm a pulsar discovery, radio and/or gamma-ray timing \citep{Ray2011} allows the gamma rays to be phase-folded
over a longer epoch, and the pulsed significance becomes high enough that trials penalities incurred by optimizing $\mu_w$ do not compromise
confidence in the detection. 

PSR B1509$-$58 has long been notorious for having by far the softest spectrum of any gamma-ray pulsar, with a spectral cut-off below 50 MeV \citep{LAT_PSR_B1509-58}.
It is so bright that it is easily detected for a broad range of $\mu_w$ values, but its highest pulsed significance is with the low value of $\mu_w = 2.3$.
\citet{KuiperHermsen2015} argue that PSR B1509$-$58 is one of a class of pulsars with spectra peaking in the MeV range. 
To enhance our sensitivity to faint soft-spectra pulsars, we repeated our search with $\mu_w = 2.3$. 
%Similarly, to search for hard-spectra pulsars in confused regions, we also tried $\mu_w$ values of $4.2$ and $4.6$.
%No additional gamma-ray pulsars were discovered.
%
%We note that all 16 pulsars are detected using photons with $E_\gamma > 100$ MeV and $\Delta\theta < 2^\circ$.
%The data files are $\approx 6$ times smaller, easing calculations.
% BELOW IS EMAIL SENT TO PHB, above is the old two lines.
No new pulsars were found.
We note that the gamma-ray pulsar detections described below were initially made using photons with  $E_\gamma > 100$ MeV and $\Delta\theta < 2^\circ$.
The data files are $\approx 6$ times smaller than the $E_\gamma > 50$ MeV data sets, easing calculations, albeit with less sensitivity to pulsars having spectra like PSR B1509-58's.

%We first folded the thousand pulsars using photons with $E_\gamma > 100$ MeV, from within $\Delta\theta < 2^\circ$ of the pulsar position. 
%The values of $\mu_w = 3.2,\, 3.6$ and $4.0$ flagged all sixteen gamma-pulsation discoveries reported below.
%Concerned, however, that both the data sample and the weighting parameters are biased against faint soft-spectra pulsars,
%we adopted the $E_\gamma > 50$ MeV, $\Delta\theta <5^\circ$ selection described in Section \ref{WeightSection} and repeated the study using $\mu_w$ values of $2.3$ and $2.8$.
%We also scanned $\mu_w$ values of $4.2$ and $4.6$, in search of hard-spectra pulsars in confused regions.
%No additional gamma-ray pulsars were discovered. 
%More of the gamma-ray pulsars had somewhat improved significance than the contrary, so we kept the $E_\gamma > 50$ MeV, $<5^\circ$ data sample.
%\citet{SearchPulsation} pushes beyond Eq. \ref{weightEq} to further improve LAT's sensitivity to faint pulsars, with results to be presented elsewhere.
%Preliminary results indicate two additional pulsar discoveries, 
%below detection threshold with the methods described here\footnote{Philippe, if you present the
%discovery of B0114+58 and B0136+57, you'll need to invite the JBO gang to sign your paper.}.

\subsection{Rotation Ephemerides}
Phase-folding gamma rays requires phase-connected rotation ephemerides accurate during the {\em Fermi} epoch.
Ephemeris timing residuals with root-mean-squares smaller than $P/50$ provide adequate accuracy for all gamma-ray pulsars known so far.
Routine observations of many pulsars stop within a few years of their discovery, however, and such ephemerides often do not exist. 
For this reason, to prepare for the {\em Fermi} mission, radio and X-ray astronomers organized to provide
ephemerides for the 230 pulsars with spindown power $\dot E > 10^{34}$ erg s$^{-1}$ that were known at the time \citep{TimingForFermi}.
Astronomers from Parkes Observatory in Australia \citep{ParkesFermiTiming}, from Jodrell Bank Observatory in England \citep{Jodrell}, 
and from Nan\c cay in France \citep{Cognard2011} continue to provide timing data not only for most of the original list, 
but also for high $\dot E$ pulsars discovered subsequently and, key to this work, for a thousand pulsars with lower spindown power. 
Astronomers from Effelsberg (Germany), Green Bank (USA), Arecibo (Puerto Rico), the Giant Metrewave Radio Telescope (India),  Westerbork (Netherlands), 
and other radio telescopes, as well as X-ray astronomers also 
help extract as many pulsed signals as possible from the LAT data by providing rotation ephemerides to the LAT team. 
In particular, the Green Bank Northern Celestial Cap (GBNCC) pulsar survey \citep[][and references therein]{GBNCC-III} and 
the High Time Resolution Universe (HTRU) pulsar survey \citep[][Cameron et al. in prep]{Ng2015_HTRUN, Cameron2018_HTRUN} provided dozens of rotation ephemerides.
%3PC will detail the timing used for the LAT pulsar discoveries.
Half of the $\sim 2800$ known radio pulsars\footnote{See the ATNF Pulsar Catalog, \url{http://www.atnf.csiro.au/research/pulsar/psrcat}} \citep{ATNFcatalog}
have not been explored for gamma-ray pulsations. 
Attempts to use older rotation parameters to seed a focused periodicity search engender so many trials
that only high-significance detections would be compelling, and are beyond the scope of this work. 

We exclude the $\sim 170$ known pulsars in globular clusters from the results shown here: 
their unknown orbital acceleration within the cluster generally skews their observed $\dot E$ values relative to their true spindown powers,
diminishing their usefulness for this study.
We have however gamma ray phase-folded 53 for which we could obtain reliable ephemerides, using the methods described here.
Beyond the re-detection of the two reported in 2PC, no new discoveries were made.

``Ephemeris validity'' strictly speaking is the interval between the first and last time-of-arrival measurements used to determine the rotation spindown parameters.
However, for many pulsars spindown is stable and the predicted number of neutron star turns is accurate for months, even years, before and/or after that interval. 
Conversely, young high $\dot E$ pulsars can have erratic spindown fluctuations (``timing noise'') requiring rotation models to use many fitted parameters.
MSPs can too, although the fluctuations are  generally smaller \citep{MSPtimingNoise}. 
Binary MSPs can undergo erratic orbital perturbations when the pulsar wind ablates the companion, 
as for example for PSR J1048+2339 \citep{Deneva_J1048}.
Models for such pulsars predict the number of rotations, and thus phase, poorly outside the validity interval.
%{\em Fermi} LAT's sky survey accumulates exposure uniformly on every part of the sky.
%Combined with pulsars' steady gamma-ray emission and constant background rates, 
%the pulsed H-test of a signal increases linearly with time for as long as the rotation model is accurate.
%The new gamma-ray pulsars discussed below all show steady significance growth over the entire {\em Fermi} mission regardless of the ephemeris validity,
%with the exceptions of PSRs J1327$-$0755 and J1925+1720, for which no signal accumulates before the validity start.

\begin{figure}[ht!]
\centering
\includegraphics[width=0.48\textwidth, angle=0]{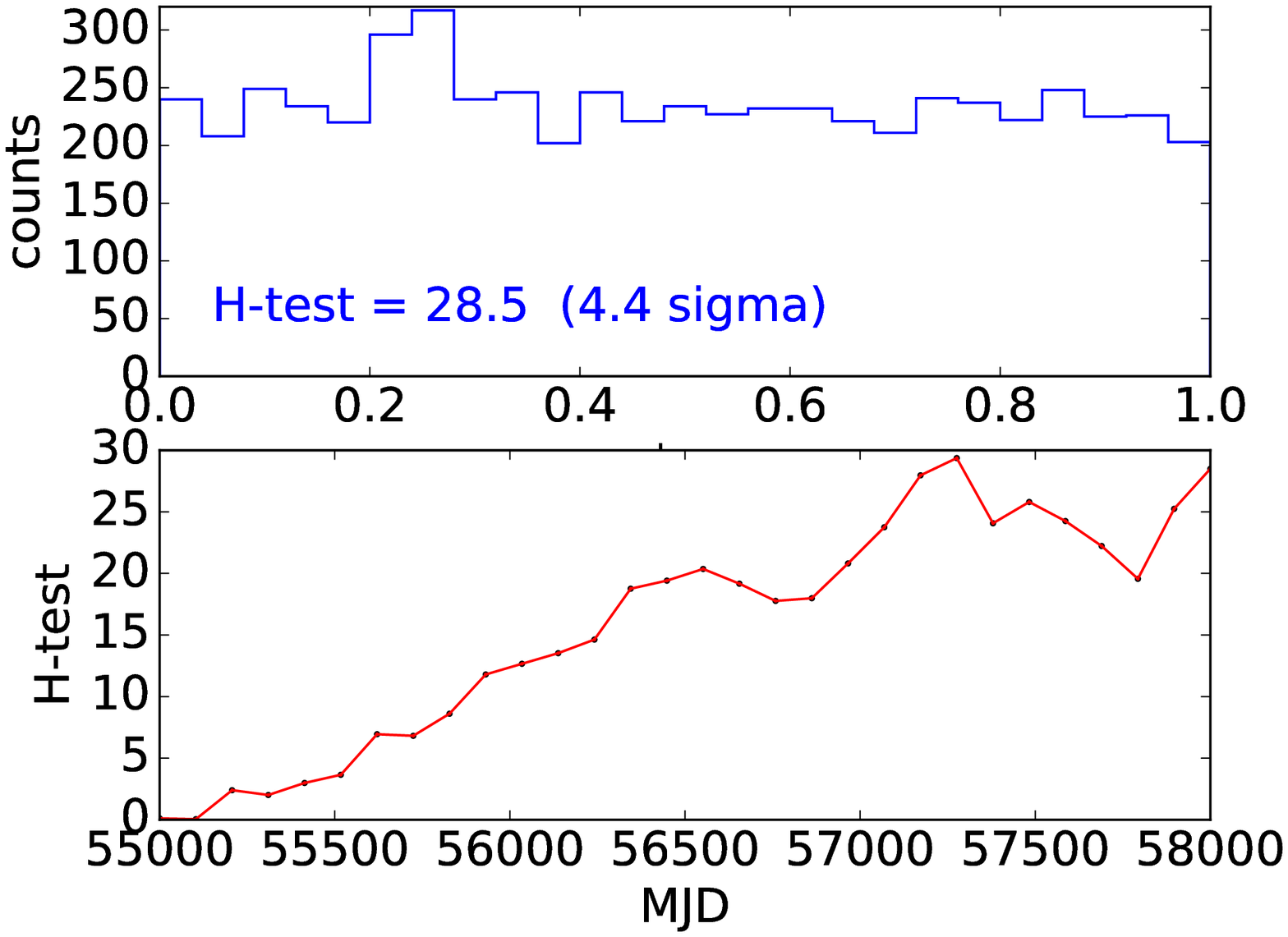}
\includegraphics[width=0.48\textwidth, angle=0]{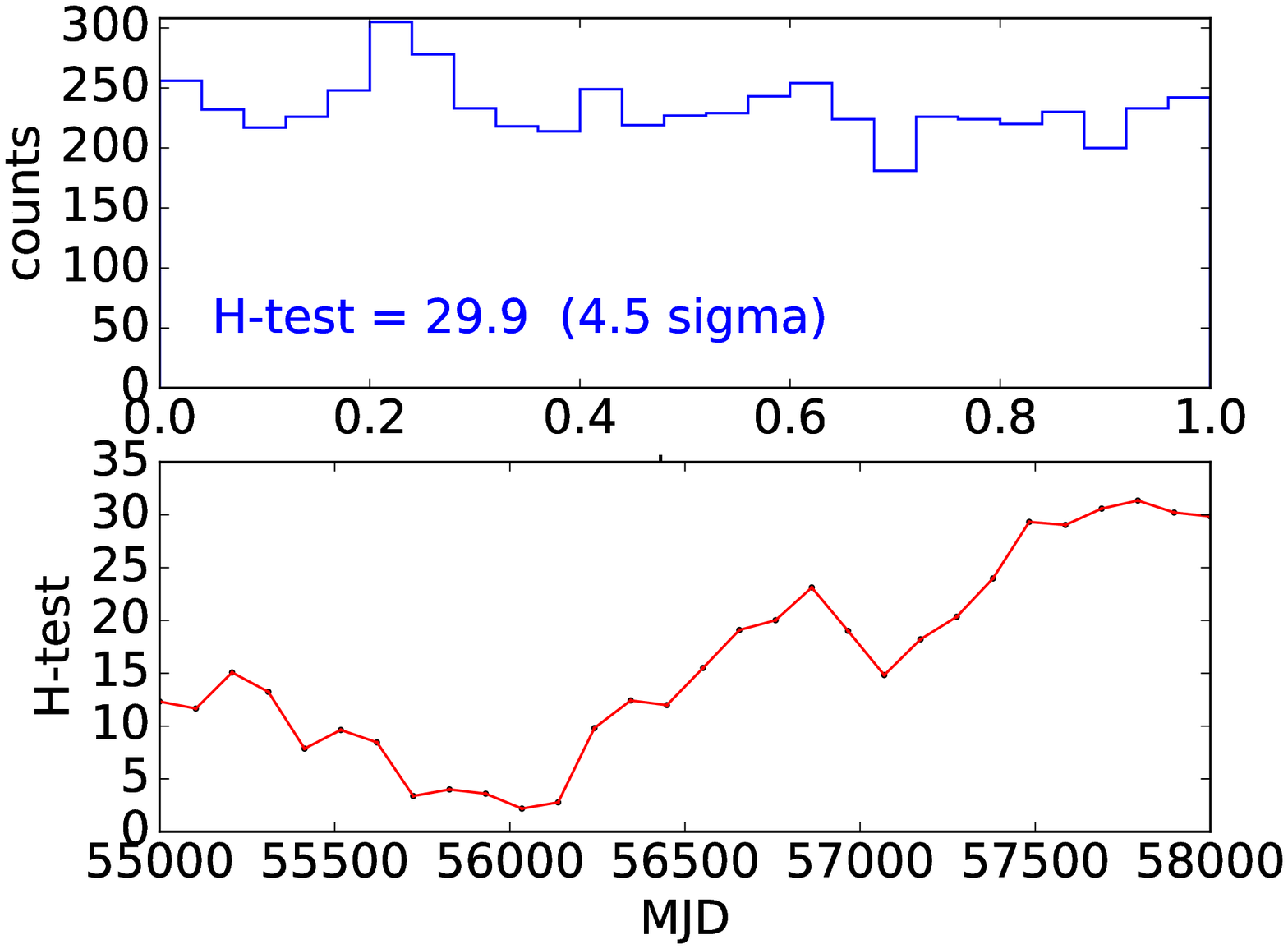}
\includegraphics[width=0.48\textwidth, angle=0]{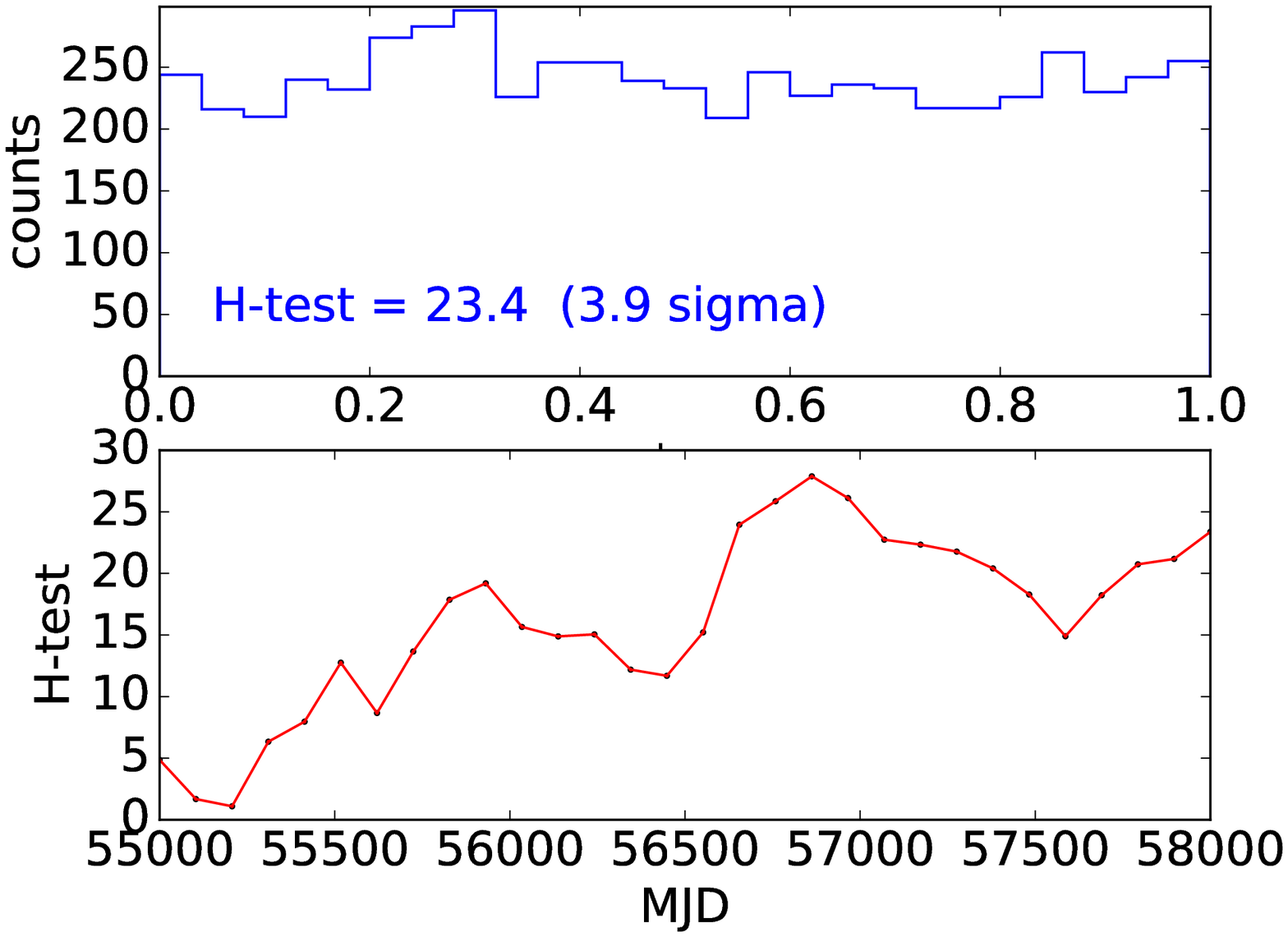}
\includegraphics[width=0.48\textwidth, angle=0]{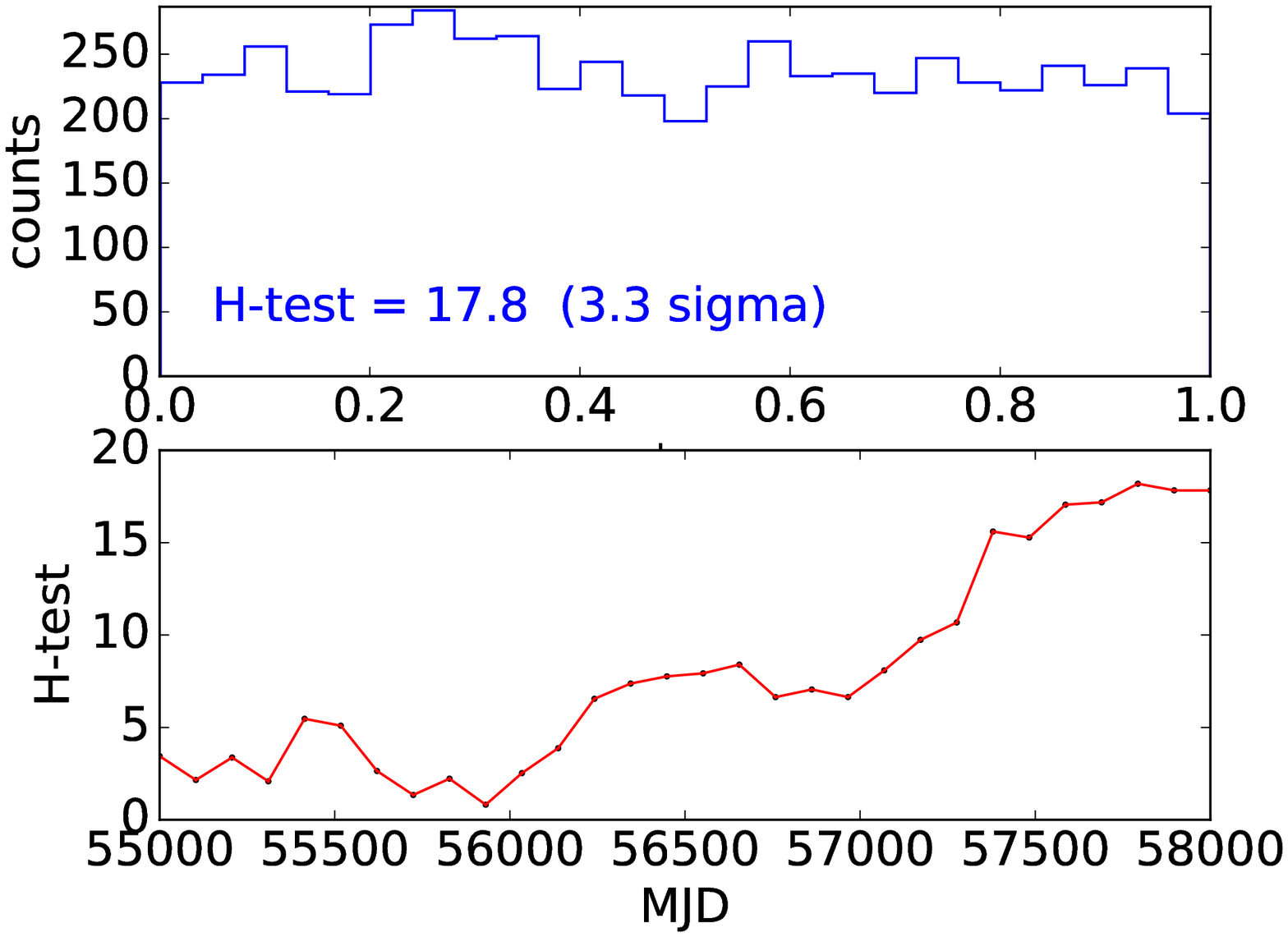}
\caption{Four simulated LAT 8.2-year (3000 day) missions for a single pulsar with flux $\sim 1 \times 10^{-9}$ photons cm$^{-2}$ s$^{-1}$
and a typical low background rate. 
Top frames: phase-histogram integrated over the mission.
Bottom frames: Integrated significance as the data accumulates over time (MJD = Modified Julian Day).
The pulsar is detected when H-test $>25$. For these parameters, half of the trials yield H-test $<15$.
Large fluctuations in cumulated significance result from Poisson fluctuations in the low photon arrival rate.
\label{simHvsMJD}}
\end{figure}

%\begin{figure}[ht!]
%\centering
%\includegraphics[width=0.43\textwidth, angle=0]{HvsMJDmc_0.10_100_0.25_hist.eps}
%\includegraphics[width=0.43\textwidth, angle=0]{SigVsFlux.eps}
%\caption{A simulated square gamma-ray pulse $0.1$ rotations wide and
%the low background level of a high latitude pulsar.
%Left: 
%For flux  $\sim 1.2 \times 10^{-9}$ photons cm$^{-2}$ s$^{-1}$,
%the distribution of unweighted H-test significances (sigma units) for 300 simulations of a 3000 day mission.
%Photon position smearing is neglected.
%Right: 
%Pulsation significance  versus the simulated flux (solid curve).
%The dashed curves show the 95 percentiles.
%Horizontal dashed lines show the 4$\sigma$ and 5$\sigma$ 
%significance thresholds. 
%The lowest fluxes reported in 2PC were near $10^{-9}$ photons cm$^{-2}$ s$^{-1}$,
%indicated by a vertical dotted line.
%I'll get rid of the other vertical line.
%\label{SigVsFlux}}
%\end{figure}

\section{Searching a Thousand Pulsars for Gamma Rays}  \label{FoldingSection}
In practice, we define $H_\mathrm{max}$ as the maximum H-test value for 6 trials: the 3 values of $\mu_w = 3.2,\, 3.6$ and $4.0$ described above, 
applied to two data samples. One sample uses $9.6$ years of LAT data, from the beginning of the {\em Fermi} mission, MJD 54682, until MJD 58182. 
The other uses LAT data only during the ephemeris validity period.
Many pulsars rise to higher significance followed by a rollover at intermediate dates, 
but restricting to two epoch durations keeps the number of trials small and well-defined.

Figure \ref{H36_EDOT} shows $H_\mathrm{max}$ versus $\dot E$ for 1269 pulsars, and
%The 15 new gamma-ray pulsars discovered during this search are discussed below.
Figure \ref{HistoHs} shows a histogram of the same $H_\mathrm{max}$ values. 
A ``noise floor'' appears. 
We choose $H_\mathrm{max} > 25$ (single trial weighted H-test significance of $4.1\sigma$) as our threshold for detection of gamma-ray pulsations. 
With this threshold, we expect less than one false detection for our 1269 pulsar sample.
We discuss the 16 gamma-ray pulsars thus discovered below.

Figure \ref{HistoHs} also shows the distribution of H-test values obtained for the same pulsar sample,
where the photon phases were replaced with uniform random values between 0 and 1, using  $\mu_w = 3.6$.
We did this 3 times (solid histogram) and 8 times (dotted histogram) for each pulsar and kept the maximum value, as in the procedure for $H_\mathrm{max}$.
The 3-trial histogram matches that of the undetected pulsars.
Three trials suffice because the $\mu_w = 2.3$ trials are not included in Figures \ref{H36_EDOT}  and \ref{HistoHs}, 
but also because counting both data sets is redundant: the ``all data'' and ``ephemeris validity'' H-test values are in fact highly correlated.
Two-thirds (one-third) of our ephemerides cover over half (over 90\%) of the mission. 
Ephemeris validity is $<500$ days for only 5\% of the sample.
Eight trials adds the fourth scan, with $\mu_w=2.3$, for the two data samples, ignoring correlations between the samples.
$H_\mathrm{max} > 25$ never occurs in the random histograms.

A simple Monte Carlo calculation, illustrated in Figure \ref{simHvsMJD}, gives insight.
Four simulated examples of significance growth versus time are shown.
For the few photons per month signal rates evoked in the Introduction for pulsars near detection threshold,
Poisson fluctuations can cause irregular growth, sometimes wildly so.
We see such fluctuations in the data, and speculated that they could be due to changes in flux, instrument exposure,
or ephemeris instability. In fact, Poisson fluctuations dominate.
%The simulations clearly confirm that $H_\mathrm{max} > 25$ ($4.1\sigma$) avoids false positive detections.  
For the simulated signal and background rates in Figure \ref{simHvsMJD}, $F_{100}^\mathrm{min} \simeq 10^{-9}$ photons cm$^{-2}$ s$^{-1}$, 
we simulated the {\em Fermi} mission 1000 times. The mean single-trial significance is $\sim 4\sigma$,
meaning that the pulsar is not detected half the time. 
The full width at half maximum of the significance distribution extends from  $2.5\sigma$ to $5\sigma$.
%Repeating the study for a range of signal strengths shows that $4.1\sigma$ threshold
%yields roughly 20\% more detections than the $5\sigma$ threshold we had applied in previous studies.
%The $H_\mathrm{max}$ procedure we use became possible with the availability of the simple weighting method,
%and in turn allows the well-defined number of trials that makes the false positive rate negligible with the $4.1\sigma$ threshold.
%
We also used the simulation to understand how many years it takes to detect a pulsar having a
given flux, duty cycle, and background environment. Statistical fluctuations smear the answer.
Near threshold, the fluctuations are Poissonian and can be so large that % for many simulated missions,
$H_\mathrm{max} > 25$ can occur a full year earlier or later than the average duration.

%Of the 580 pulsars with  $\dot E < 10^{33}$ erg s$^{-1}$,
%four have $ H_\mathrm{max} > 15$, which corresponds to $3\sigma$ significance. One is PSR J2208+4056, whose detection is discussed below.
%The other three values are 16, 17, and 21.
%Of the high $\dot E$ undetected pulsars, 7 (4) have $H_\mathrm{max} > 15 \, (20)$, respectively, but $ H_\mathrm{max} < 25$.
%Some of the latter will likely exceed detection threshold in the near future.
%
%Another interesting consequence is that detections sometimes ``pop up'' sooner than
%a linear extrapolation of an observed $H_\mathrm{max}$ versus MJD curve predicts,
%due to a positive burst of events. Of course, negative fluctuations can make
%a pulsar take longer than hoped to appear.

% Diffuse flux is 2 to 15 x 1.E-4 gammas/sr/cm²/s for most of the pulsars in this paper,
% that is, l=-18 to l=+57 degrees, and smack on the plane.
% Diffuse Gamma-Ray Emission in the Galactic Plane from Cosmic-Ray, Matter, and Photon Interactions
% Bertsch, D. L.; Dame, T. M.; Fichtel, C. E.; Hunter, S. D.; Sreekumar, P.; Stacy, J. G.; Thaddeus, P.
% http://adsabs.harvard.edu/abs/1993ApJ...416..587B
% 1 degree radius is 1/60 rad, or pi/60/60 = 0.00087 sr.
% Diffuse flux in a year is 3.14E7*1.E-4*8000*0.00087/6  where 6 is for the duty cycle.
% This gives 3600 photons per year, divided into 25 phase bins ==> 144/bin.

\begin{figure}[ht!]
\centering
\includegraphics[width=1.1\textwidth]{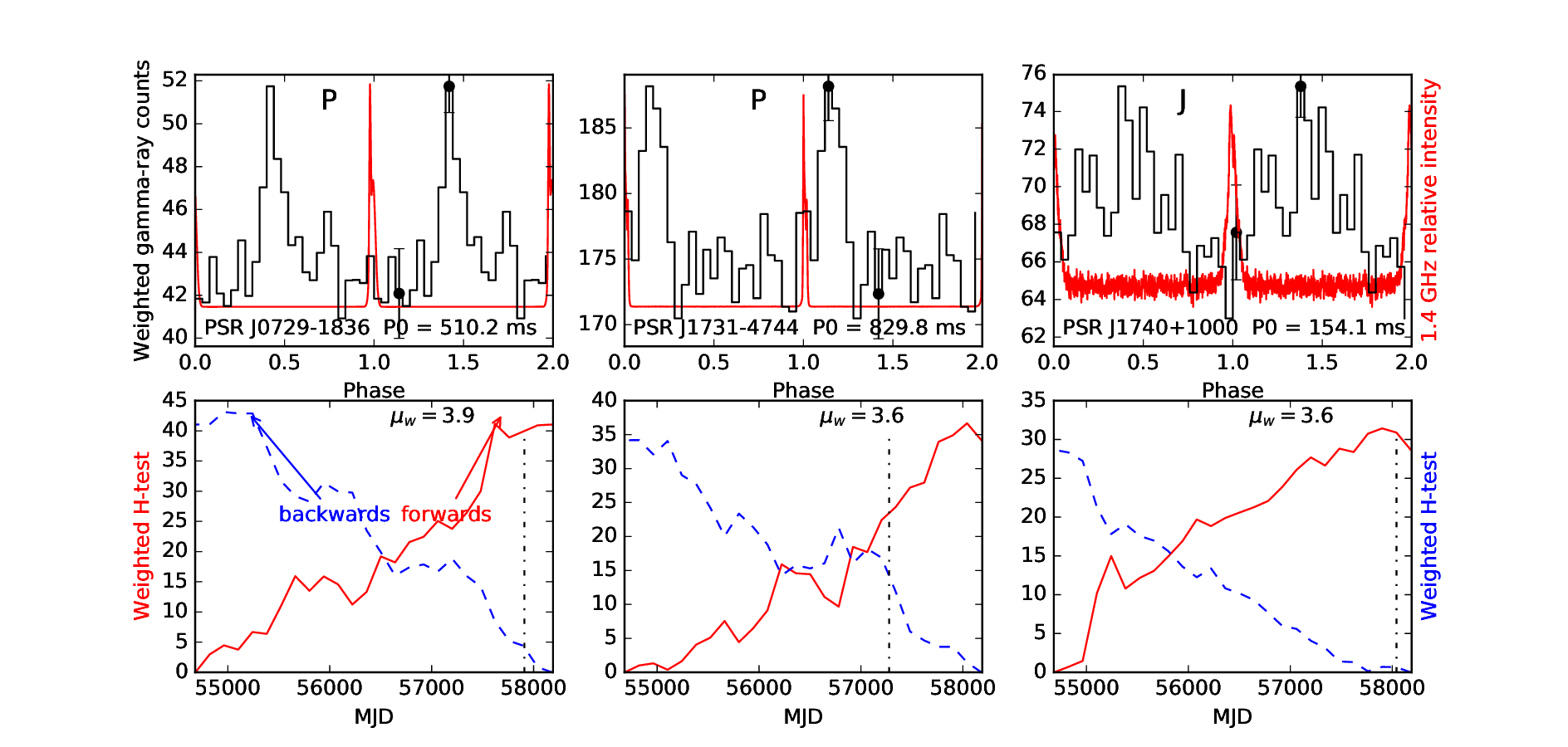}
\includegraphics[width=1.1\textwidth]{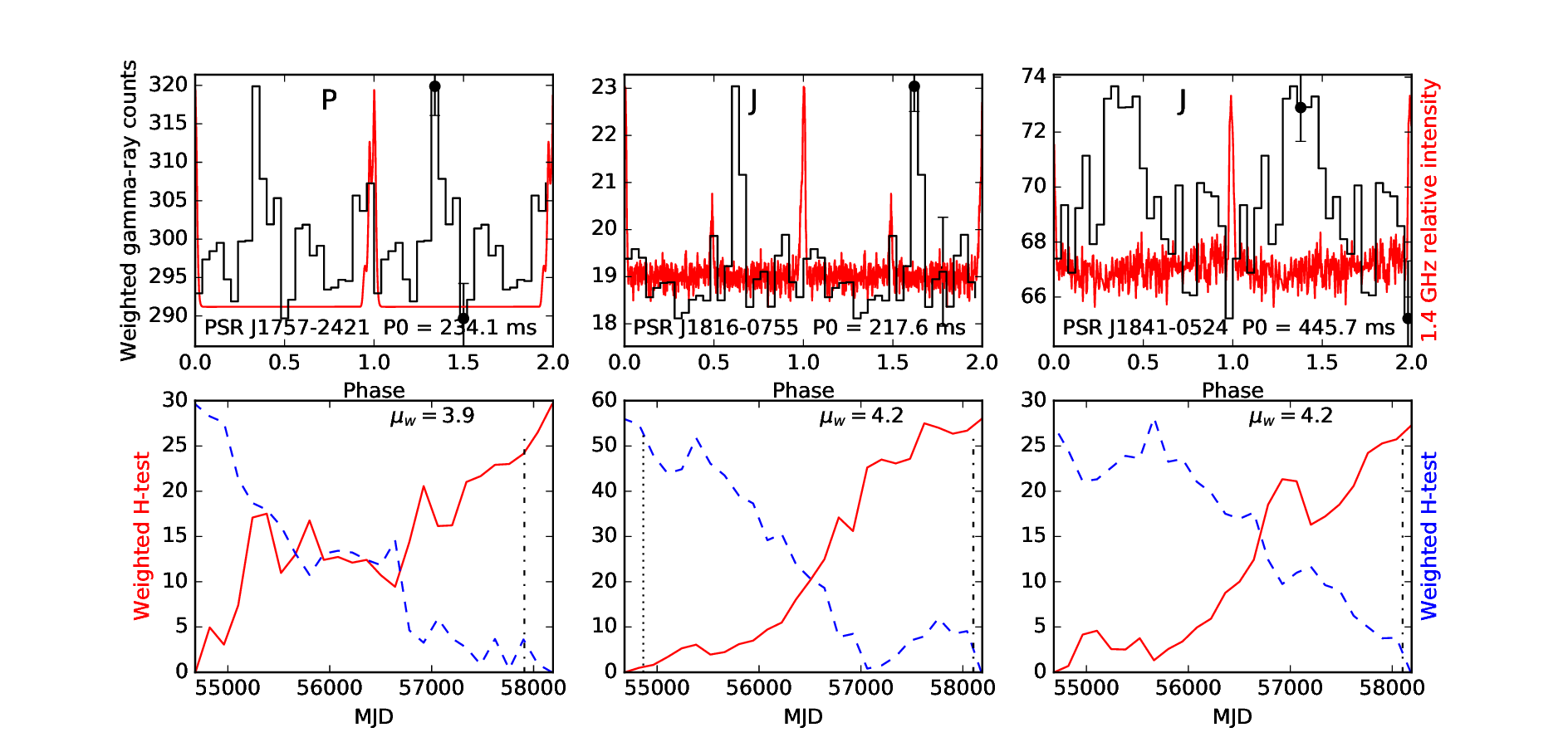}
\caption{Six young and middle-aged pulsars newly discovered in gamma rays.
Top: Weighted phase histograms of the $>50$ MeV gamma rays. 
The largest and smallest statistical uncertainties are shown.
The overlaid $1.4$ GHz radio profiles are phase-aligned, with 
the letter indicating the radio telescope that recorded the profile: N, Nan\c cay; P, Parkes ; J, Jodrell Bank,
see the references in Table \ref{tbl-charPSR}.
A full rotation is shown twice, with 25 phase bins.
Bottom: H-test significance cumulated starting at both the beginning (solid curve) and at the end (dashed
curve) of the dataset. The vertical dotted (dot-dash) line indicates the start (finish) of the ephemeris validity.
The weighting parameter $\mu_w$ used for each pulsar is shown.
% The phase uncertainty due the DM precision only shows up for J1125, Dave look into the others.
\label{YoungPSRs1}}
\end{figure}

\begin{figure}[ht!]
\centering
\includegraphics[width=1.1\textwidth]{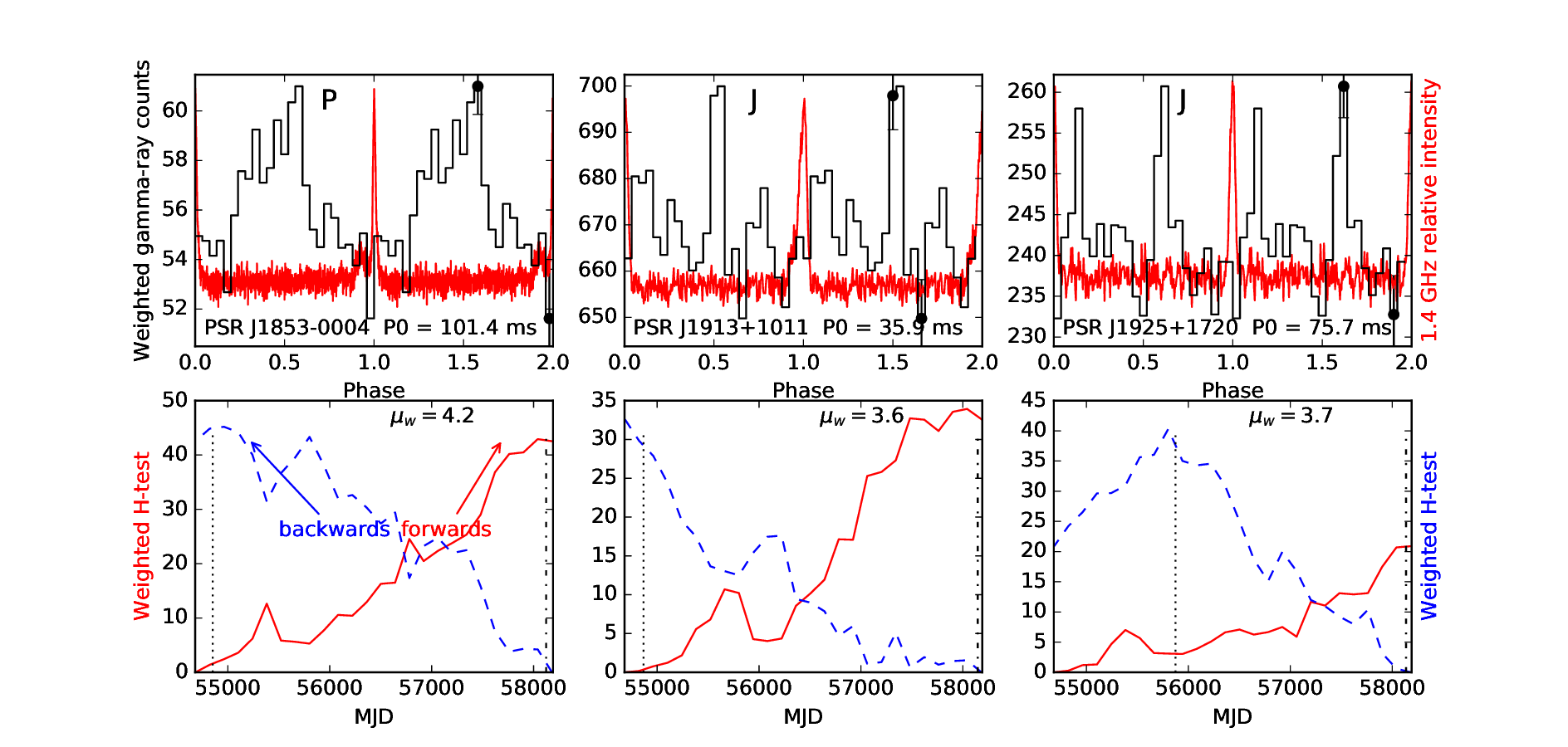}
\includegraphics[width=1.1\textwidth]{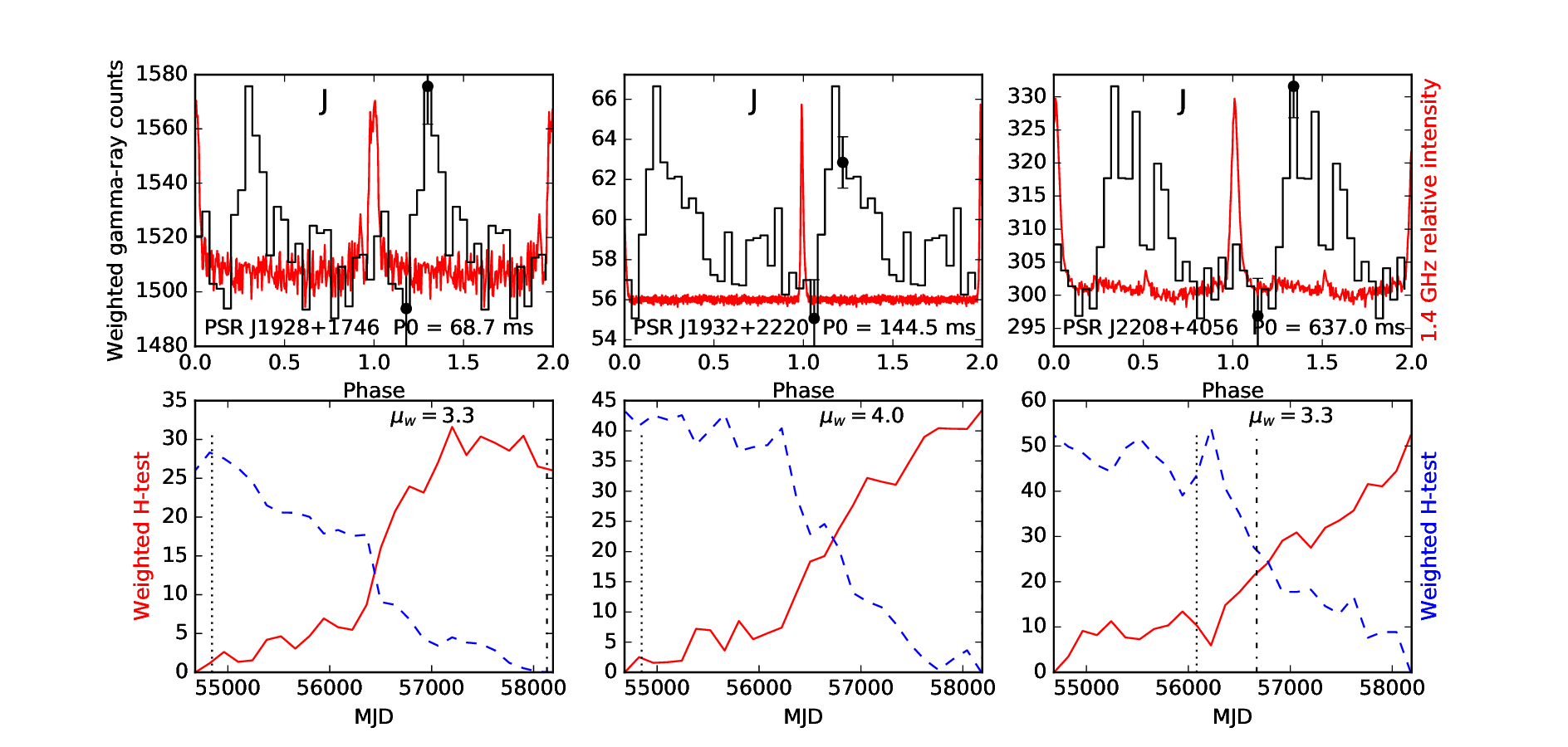}
\caption{Six young pulsars newly discovered in gamma rays, as in Figure \ref{YoungPSRs1}.
For PSR J1925+1720 the phase histogram integrates data only after the beginning of the ephemeris validity. 
For the other pulsars, all data from MJD 54682 until MJD 58182 is used. 
\label{YoungPSRs2}}
\end{figure}

\subsection{Discoveries of Gamma Rays from Radio Pulsars} \label{NewPulsarsSection}
%\subsection{Radio Pulsars Newly Detected in Gamma-rays} \label{NewPulsarsSection}
%\subsection{New Gamma-ray Pulsars} \label{NewPulsarsSection}
%
Table \ref{tbl-charPSR} lists 16 gamma-ray pulsars discovered using the above method.
Twelve are young to middle-aged, and four are MSPs. 
Figures \ref{YoungPSRs1}, \ref{YoungPSRs2} and \ref{MSPs3} show their phase-aligned $>50$ MeV gamma-ray and radio profiles,
as well as the evolution of their H-test (pulsed Test Significance) since the start of the {\em Fermi} mission. 

\subsubsection{Non-recycled Pulsars}
The parameter space occupied by gamma-ray pulsars continues to fill in and expand. 
PSR J2208+4056 has $\dot E = 8\times 10^{32}$ erg~s$^{-1}$, 
three times lower than the previous lowest spindown power for young gamma-ray pulsars.
PSR J0729$-$1836 is now one of nine young pulsars with $\dot E < 10^{34}$ erg s$^{-1}$.
At the other end of the range, PSRs J1913+1011 and J1928+1746 both have $\dot E > 10^{36}$ erg s$^{-1}$.

PSR J1731$-$4744 is the longest spin-period gamma-ray pulsar now known, with $P = 830$ ms.
Slow pulsars (PSR J0729$-$1836 has $P = 510$ ms) attain $\dot E \simeq 10^{34}$ erg s$^{-1}$
because their surface magnetic fields are high, $B_S \ga 10^{13}$ G.
In contrast, PSR J1913+1011 is nearly as fast as the Crab but with $B_S$ an order of magnitude smaller.

%\cite{SixWeak} define gamma-ray pulsar ``\textit{faintness}'' as stemming not just from the pulsars' luminosity or distance, 
%but also from the pulsed duty cycle, the background level, and the spectral shape. 
In the LAT 8-year preliminary source list\textsuperscript{\ref{fsscfootnote}}, FL8Y, 
only the six pulsars flagged in Table \ref{tbl-charPSR} lie within a source's 95\% confidence ellipse.
The rest are undetected, using a procedure similar to that of 3FGL \citep{3FGL}, improved in preparation for 4FGL, the upcoming $4^{th}$ {\em Fermi} LAT catalog.
Weighted folding improves signal-to-noise roughly as the inverse of the pulse width $\delta\phi$,
compared to a phase-integrated detection.
Seven of the twelve young pulsars have Galactic latitude $|b| < 0^\circ .6$ (and thus very high background levels),
and all but one of these (PSR J1853$-$0004) have narrow gamma-ray pulses, with $\delta\phi < 0.3$ in phase.
%The only broad pulse in Figure \ref{YoungPSRs}
%is for PSR J1740+1000 which is, in turn, the only young pulsar at high latitude, $b = 20^\circ$, with
%few nearby sources, and thus with low background.

The highest background levels are towards the Galactic center.
The detection of PSR J1757$-$2421, only  $5^\circ .3$ away, is aided by its narrow pulse, $\delta\phi <0.1$. 
%The LAT's broad PSF also makes pulsars hard to detect in populated regions. 
PSR J1841$-$0524, with high $\dot E$ and intermediate distance, presumably needed over nine
years of LAT data to be seen because its overlap with the bright pulsar wind nebula HESS J1841$-$055 raises its background \citep{Rousseau2013}.
Only $0^\circ .9$ separate  high $\dot E$ PSRs J1925+1720 and J1928+1746, raising each's background marginally compared to the already
high level of this busy region in the Galactic plane, compounding detection difficulty.

Finally, PSR J1932+2220 is the most distant of the new pulsars. 
The NE2001 \citep{Cordes2002} and the YMW16 \citep{ymw16} Dispersion Measure (DM) models place it at $7.5$ and 8 kpc from Earth, respectively,
whereas its distance derived from HI absorption is $10.9^{+0.3}_{-0.8}$ kpc \citep{Verbiest12}.
The others pulsars lie in the range of $\sim 1$ to $\sim 6$ kpc, like the bulk of the 2PC pulsars. 
PSR J1932+2220's ``faintness'' is thus a combination of low flux (medium $\dot E$ at large distance) and 
the background level just off the plane.

\subsubsection{Millisecond Pulsars}
\begin{figure}[ht!]
\centering
\includegraphics[width=0.7\textwidth]{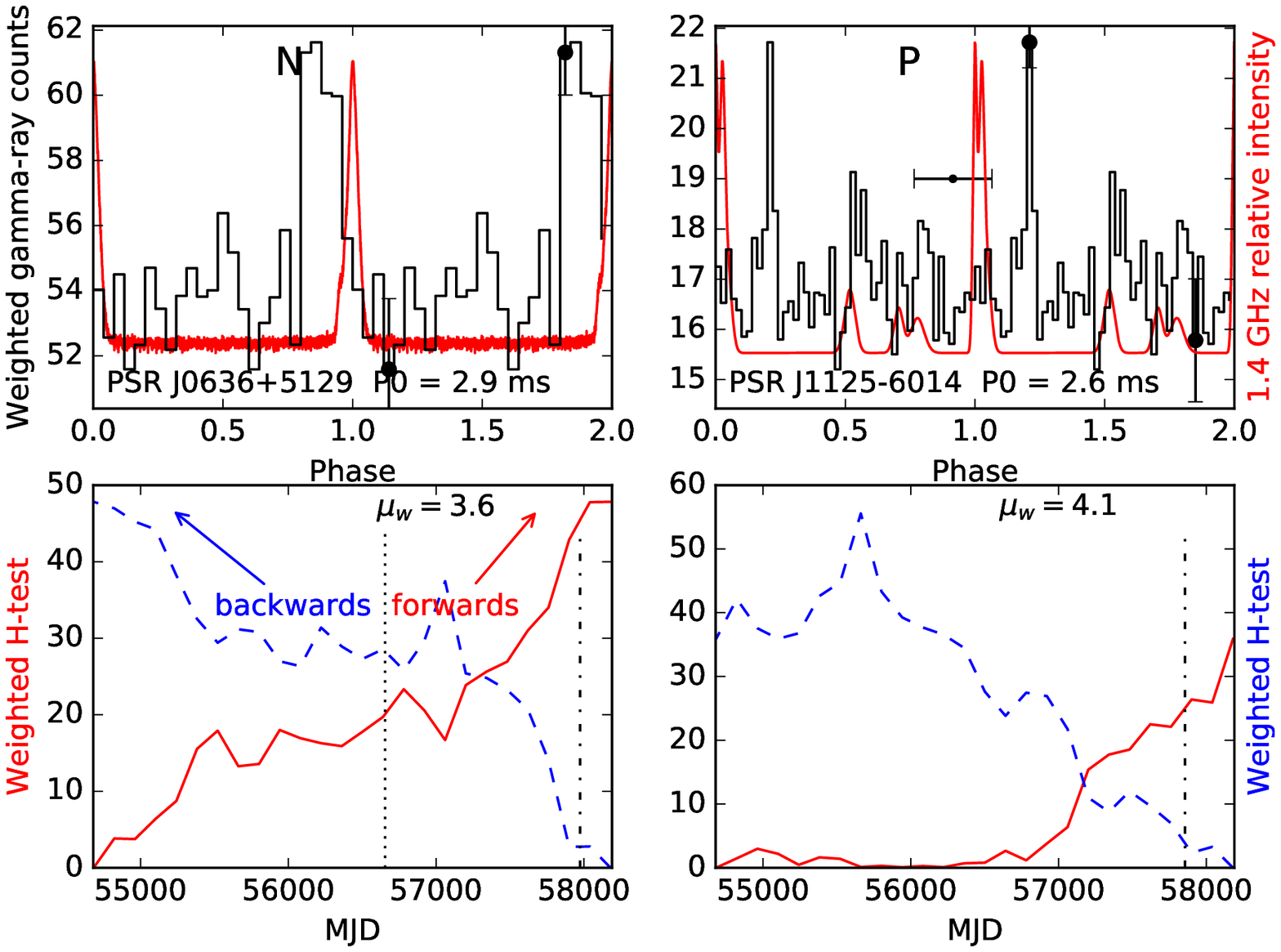}
\includegraphics[width=0.7\textwidth]{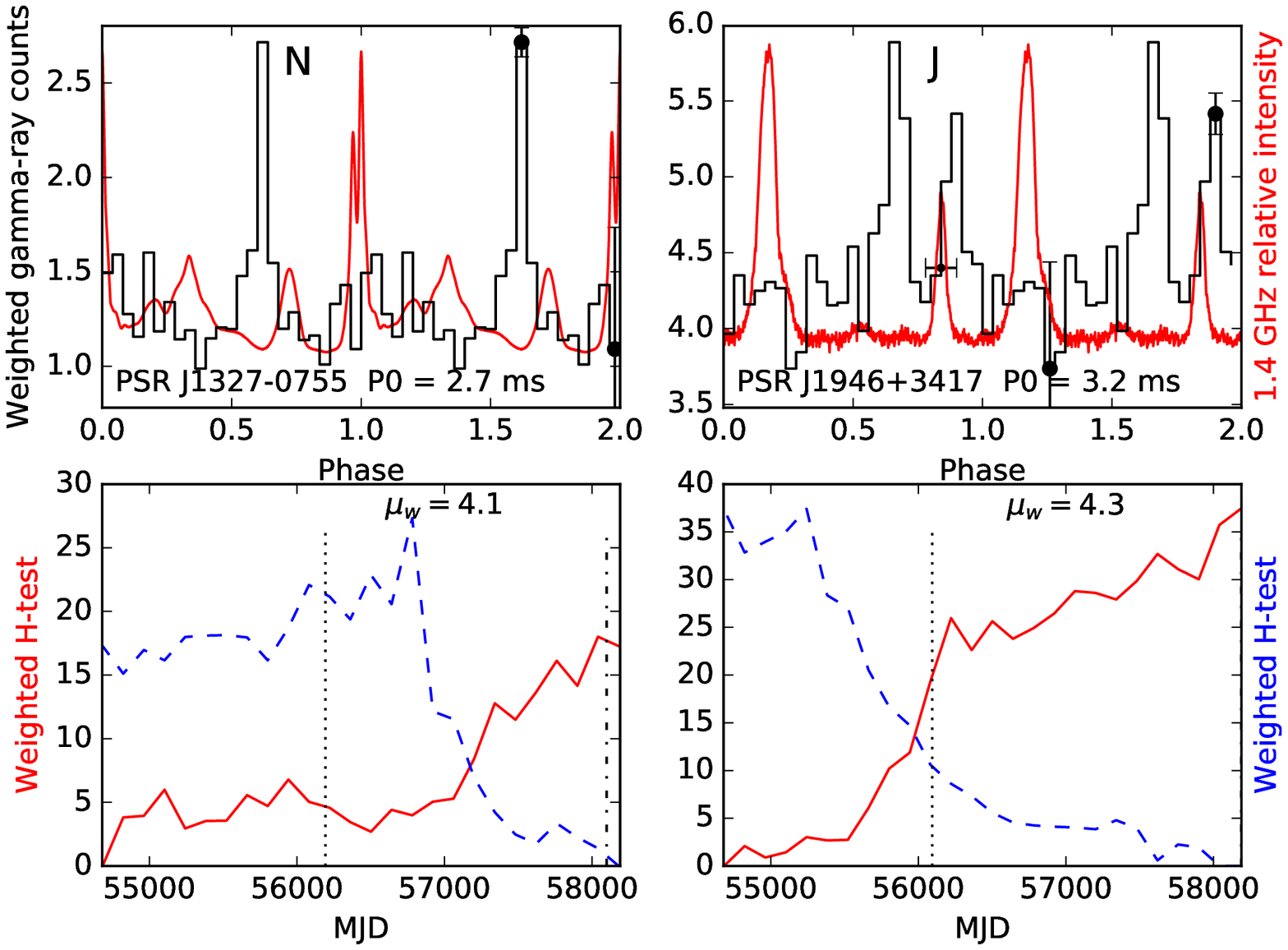}
\caption{Four MSPs newly discovered in gamma rays.
Top: Weighted phase histograms of the $>50$ MeV gamma rays. 
The largest and smallest statistical uncertainties are shown.
The overlaid $1.4$ GHz radio profiles are phase-aligned, with 
the letter indicating the radio telescope that recorded the profile: N, Nan\c cay; P, Parkes ; J, Jodrell Bank,
see the references in Table \ref{tbl-charPSR}.
The DM uncertainty leads to the alignment uncertainty shown by the horizontal error bars for two MSPs.
For the other two it is $<0.01$ and is not shown.
A full rotation is shown twice, with 25 phase bins (50 bins for J1125$-$6014).
Bottom: H-test significance cumulated starting at both the beginning (solid curve) and at the end (dashed
curve) of the dataset. The vertical dotted (dot-dash) line indicates the start (end) of the ephemeris validity.
The weighting parameter $\mu_w$ used for each pulsar is shown.
For PSR J1327$-$0755 the phase histogram integrates data only after the beginning of the ephemeris validity. 
For the other pulsars, all data from MJD 54682 until MJD 58182 are used.
\label{MSPs3}}
\end{figure}

Three of the four new MSPs are ``faint'' simply due to low gamma-ray fluxes: neither the intense background at low latitudes
nor broad peaks made them hard to detect. Three have distances $d<2$ kpc, as derived from their DM values. 
The only one in the FL8Y list, PSR J1125$-$6014, is also the only one deep in the plane, with $b = 0^\circ .9$, with five other gamma-ray pulsars
less than $2^\circ$ away.
It also has the narrowest pulse, facilitating detection. In Figure \ref{MSPs3} its signal is contained in a single
phase bin of width $\delta\phi = 1/50 = 0.02$ in phase, so its pulse is $<53\, \mu$s wide.

\citet {J1946p3417_Barr} used timing of the binary orbit and radio polarization measurements to characterize PSR J1946+3417.
They argue that the more intense radio peak (see Figure \ref{MSPs3}) could come from a polar cap, while the smaller one
would come from the outer magnetosphere. They determined the pulsar's geometry to be favorable for gamma-ray viewing from Earth,
and attributed its non-detection by the LAT to a large distance and low spindown power. 
Their reasoning overlaps that invoked for young radio interpulse pulsars in Section \ref{J2208sect}.
PSR J1946+3417 is shown by the lowest large triangle in Figure \ref{Edot_sqEDOTD2}, discussed below.

Having now detected it in gamma rays, we consider Barr et al's prediction that if indeed radio emission comes from an outer gap (OG), 
the gamma-ray emission would be aligned in phase with the second radio peak. 
The gamma-ray profile for PSR J1946+3417 in Figure \ref{MSPs3} shows two peaks separated by $0.25$ in phase.
The radio peak they attribute to the OG is between them, on the inner shoulder of the trailing gamma peak, consistent with their prediction.
The uncertainty shown is derived from the difference between the DM value in our ephemeris and that published by \citet {J1946p3417_Barr}.
A wealth of accurate observational information now exists for this pulsar, making it an interesting candidate for detailed modeling.

%It lies in the foreground of pulsar-rich NGC 3603, the most massive young star cluster in our Galaxy \citep{gammaNGC3603}. 
%The sharpest gamma-ray pulse known remains $23 \pm 11$ $\mu$s for PSR B1957+20 \citep{GuillemotBlackWidow_2012},
%while PSR J1231$-$1411 is $<90$ $\mu$s wide at half-maximum (see 2PC).

For any pulsar, requiring gamma-ray efficiency $\eta= L_\gamma/\dot E < 100$ \% reveals its maximum possible distance,
for a given assumption about its beaming fraction $f_\Omega$,
because the luminosity $L_\gamma = 4\pi d^2 f_\Omega G_{100}$ depends on the distance and on
the energy flux integrated above 100 MeV, $G_{100}$.
For MSPs, the Shklovskii effect makes the $\eta < 100$ \% constraint even stronger,
because the intrinsic (corrected) spindown power $\dot E^\mathrm{int}$ is also sensitive to distance, as illustrated in 2PC Figure 11.
We now explore the plausible distances of the four MSPs, and hence their possible $\dot E^\mathrm{int}$ values.

The Doppler correction to $\dot P$ depends on the system's proper motion transverse to the line of sight, $\mu$ \citep{Shklovskii}, 
and the acceleration due to its and the Sun's locations the Galaxy. 
From the observed spindown $\dot P$, the intrinsic value is
$\dot P^{\rm \,int} = \dot P -\dot P^{\rm \, shk}-\dot P^{\rm \, gal} \label{DoppEq} $, with
\begin{equation}
\dot P^{\rm \,shk} = \frac{1}{c}\,\mu^2 \,d\,P = k\,\left({\mu \over {\rm mas\, yr^{-1}}}\right)^2 \,\left({d \over {\rm kpc}}\right)\,\left({P \over {\rm s}}\right)
\end{equation}
and
\begin{equation}
\dot P^{\rm \,gal} = \frac{1}{c}\, \textbf{n}_{10} \cdot (\textbf{a}_{1}-\textbf{a}_{0})P
\end{equation}
where $k=2.43\times 10^{-21}$.  
The Galactic potential model of \cite{Carlberg_Innanen1987} and \cite{Kuijken_Gilmore1989}
provides the accelerations $\textbf{a}_{1}$ of the pulsar and $\textbf{a}_{0}$ of the Sun. 
Here, $\textbf{n}_{10}$ is the unit vector in the direction from the solar system to the pulsar.

%$\dot P^{\rm \,gal}$ is negligible for our four MSPs.
For PSR J0636+5129, the two terms nearly cancel and the Shklovskii correction is small, $<1$\%, as are the uncertainties, owing to a timing parallax distance measurement
and a small, precise $\mu$ value \citep{GBNCC-I}. 
PSR J1327$-$0755, in contrast, lies on a line-of-sight where converting the DM to distance is highly uncertain (see Table \ref{tbl-charPSR})
and its nominally large $\mu$ value has large uncertainties. We therefore do not correct $\dot P$ for this MSP.

For PSR J1125$-$6014 our ephemeris has $\mu =  17 \pm 0.1$ mas yr$^{-1}$.
Of the two distances in Table \ref{tbl-charPSR}, the closer one, $d = 1$ kpc,
gives a correction of $\dot E^\mathrm{int} = 0.6\dot E$. 
The 3FGL value of $G_{100}$ \citep{3FGL} and $f_\Omega=1$ then give a typical value of $\eta=30$\%. 
The farther distance implies small $f_\Omega$ to reach that range of $\eta$ values.

Since PSR J1946+3417 is undetected in FL8Y, we have no $G_{100}$ value, and thus no $\eta<100$\% distance constraint.
However, the ``heuristic'' energy flux $G_h = \sqrt{\dot E} /(4\pi d^2)$ can also flag suspicious distances.
In Figure \ref{Edot_sqEDOTD2}, $G_h$ versus $\dot E$ for our twelve hundred pulsars, MSP J1946+3417 is well below any other gamma-ray pulsar,
implying that the two DM distances in Table \ref{tbl-charPSR} are over-estimated.
To raise PSR J1946+3417 to $G_h = 8\times 10^{14}$ (erg s$^{-1})^{1/2}$ kpc$^{-2}$ requires $d = 2.5$ kpc, 
two or three times closer than the DM distances. % of the YMW16 \citep{ymw16} or NE2001 \citep{Cordes2002} DM models, respectively.

The Shklovskii correction lowers $G_h$ for PSR J1946+3417 even more.
Using $\mu = 8.64 \pm 0.25$ mas yr$^{-1}$ from \citet {J1946p3417_Barr}, 
for $d=7$ kpc, observed $\dot E = 3.8 \times 10^{33}$ erg s$^{-1}$ corrects to a record-breaking $\dot E^\mathrm{int} = 0.83 \times 10^{33}$  erg s$^{-1}$. 
For $d=5.2$ kpc, the corrected value is as small as any seen for a gamma-ray MSP, $\dot E^\mathrm{int} = 1.8 \times 10^{33}$  erg s$^{-1}$,
while $d=2.5$ kpc suggested by Figure \ref{Edot_sqEDOTD2} gives a more typical value of $\dot E^\mathrm{int} = 2.8 \times 10^{33}$ erg s$^{-1}$.
We add the $d=2.5$ kpc to Table \ref{tbl-charPSR} to underline that the gamma ray detection favors a closer distance.
%For its ecliptic latitude of $54^\circ$ the annual timing parallax at $2.5$ kpc would be 160 ns, far smaller than the 6 $\mu$s rms of the timing
%residuals of the rotation ephemeris we have used, 
%but perhaps within reach of the psr$\pi$ project to measure pulsar distances using the VLBA\footnote{\url{https://safe.nrao.edu/vlba/psrpi}}
%\citep[see for example][]{psrpi}. 
Finally, \citet{J1946p3417_Barr} determined this pulsar to have $1.828 \pm 0.022$ solar masses, making $I$, and thus $\dot E$,
30\% larger than obtained using the usual Chandrasekhar mass for the neutron star. 
This shifts how a gamma-ray flux constrains this pulsar's distance.

\subsection{The Birthplace of PSR J1731$-$4744}
Discovered at the dawn of the pulsar era \citep{EarlyMolonglo}, slowly spinning PSR J1731$-$4744 has a high magnetic field and thus a young characteristic age of 80 kyr. 
\citet{PM_B1727} compiled observations going back to 1971 and obtain a large proper motion with a small uncertainty, $146 \pm 12$ mas yr$^{-1}$.
We were unable to confirm their result with our timing data.
Nevertheless, they argued that at the DM distance of $5.6$ ($2.8$) kpc from the YMW16 (NE2001) model, their proper motion implies an exceptionally fast space velocity.
They propose instead that PSR J1731$-$4744 was born in the supernova of which RCW 114 is the remnant (also known as SNR G343.0$-$6.0). 
The SNR distance is between $0.5$ and 1 kpc.
For $0.7$ kpc, they find a more typical transverse speed of $v_\perp \simeq 500$ km s$^{-1}$.
An unmodeled electron excess along the line-of-sight would explain an over-estimated DM distance.
Their Figure 2 shows that the current pulsar position extrapolates back to the heart of RCW 114 over the system's age. 

\begin{figure}[ht!]
\centering
\includegraphics[width=0.75\textwidth, angle=270]{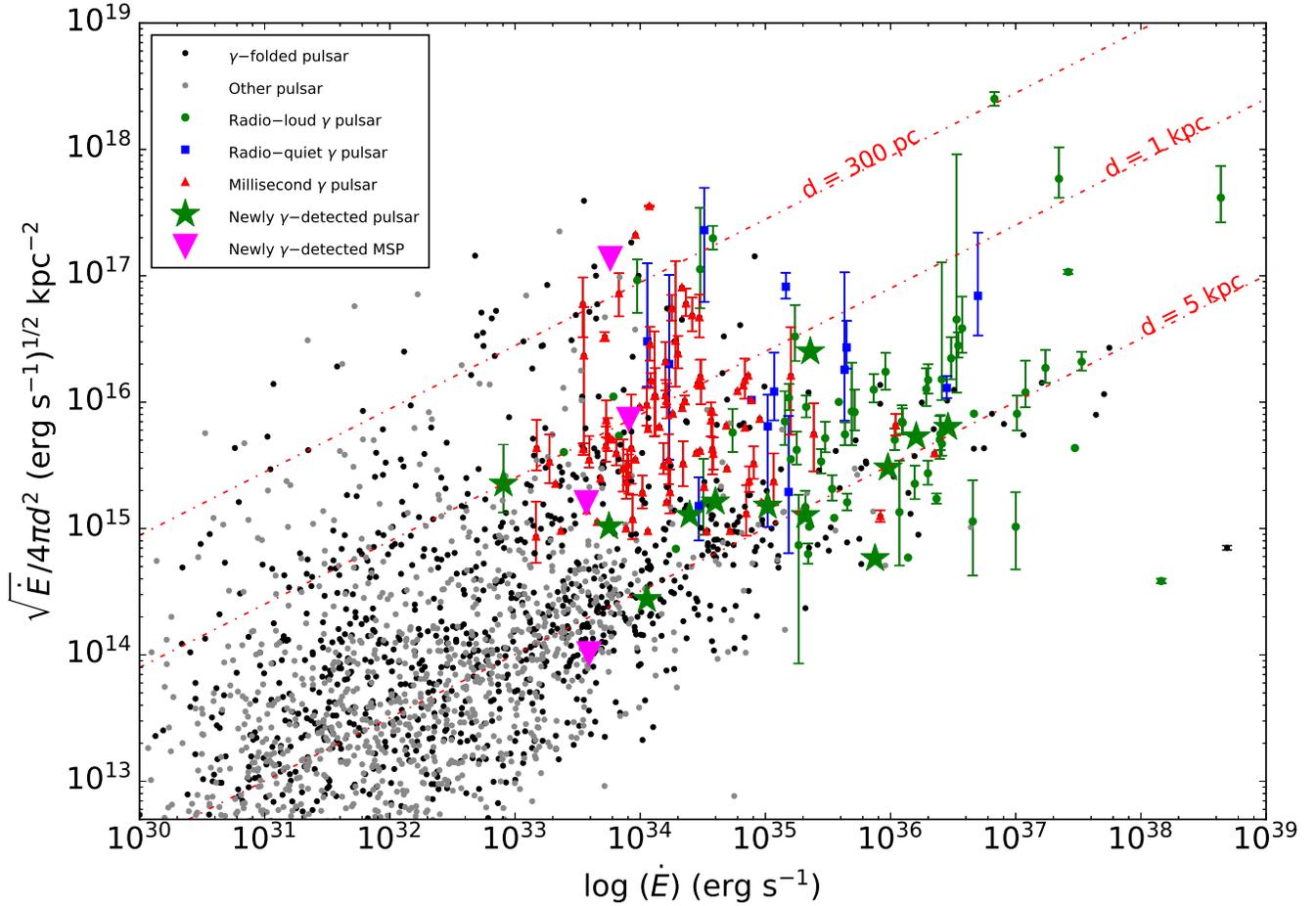}
\caption{Gamma-ray flux expected for pulsars assuming $L_\gamma = \sqrt{10^{33} \dot E} \, {\rm erg \, s}^{-1}$,
for the best distance estimate available, as a function of spindown power $\dot E$.
The diagonal dashed lines indicate constant distance. 
Grey dots indicate pulsars not gamma-folded, the other symbol codes are as in Figure \ref{H36_EDOT}.
Most nearby pulsars with $\dot E < 10^{33}$ erg s$^{-1}$ remain undetected in spite of having
heuristic fluxes two orders of magnitude larger than the measured fluxes of many pulsars with higher $\dot E$,
however the observed gamma-ray emission ``deathline'' may still be an artifact of selection bias due to the pulsar's distance.
For PSR J1327$-$0755 the second distance in Table \ref{tbl-charPSR} is used, instead of first as for the others.
To convert the figure's scale to the units of $G_{100}$, multiply by
$ 3.33\times10^{-27}\, {\rm erg}^{1/2} \,{\rm s}^{-1/2}\, {\rm kpc}^{2} \, {\rm cm}^{-2}$, see also 2PC Eq. 18. 
\label{Edot_sqEDOTD2}}
\end{figure}

The gamma-ray detection of PSR J1731$-$4744 provides an independent constraint.
The efficiency of $\eta = 11$ \% for $d = 1$ kpc is typical for 2PC pulsars (for which the spread in $\eta$ is large).
We find $\eta > 100$\% for $d >3$ kpc.
In Figure \ref{Edot_sqEDOTD2}, PSR J1731$-$4744 corresponds to the lowest star, for the YMW16 distance, even
lower than B0540$-$69 in the LMC seen at far right.

SNR G343.0$-$6.0 is not detected in the {\em Fermi} LAT catalog of supernova remnants \citep{SNRcat},
nor in the more recent search for extended Galactic sources \citep{ExtGalSources}.
Nevertheless, the gamma rays from the LAT source could be shared between the pulsar and the SNR.
%The phase-histogram shows a hint of off-pulse emission.
An extreme hypothesis that only a third of $G_{100}$ is due to the pulsar implies $\eta > 100$\% for $d >5.2$ kpc.
Comparing the on- and off-pulse spectral shapes may clarify the pulsar's contribution to the flux, although the source's faintness
will limit the information that can be extracted.
Smaller $f_\Omega$ would also enable a larger distance.
With these caveats, we conclude that the gamma-ray detection supports the association of PSR J1731$-$4744 with RCW 114,
and we adopt the nearer distance in Table \ref{tbl-charPSR}.

% Index       name         angSep    gglon     gglat    (maj,min95)  ts      nickname   assoc4FGL    target_name
% 3826  FL8Y J1730.5-4737  0.236   342.5649   -7.4337  (0.31, 0.25)  40.2  S8008-3010   G343.0-06.0   J1731-4744      edge (=yes)
% G100 is 1.O4E-11 erg/cm2/s, +/- 0.18.
%So at 1 kpc eff is 11\% (reasonable). Becomes 100\% at 3 kpc so, ignoring beaming, the NE2001 distance is at the hairy edge of plausibility,
%and the YMW16 distance is nfg. If some of the gammas come from SNR G343.0$-$6.0 then it could be farther.
%Is in LAT SNR Catalog Table 3, SNRs Not Detected by the LAT. They cite 1 kpc reference, else nothing. Busy region, of course.
%Spectral shape will be interesting in this case.
% python /work/TEMPO/DavesStuff/lumin.py
% Luminosity:  1.24431828977e+33  +/-  2.15362780922e+32
% Efficiency (%) =  11.0116662811
% hunnerd pursent distance (parsecs):  3013.51584548

%\include{IndividualPulsarStories}

%The values in parentheses are $F_{100}$, the photon flux above 100 MeV, in units of ph/s/cm2, and $G_{100}$, the energy flux integrated above 100 MeV, in units of erg/s/cm2.
%J1125-6014  (3.789e-08	1.480e-11)
%J1731-4744  (3.668e-08	1.038e-11)
%J1740+1000  (2.627e-08	7.952e-12)
%J1853-0004  (7.494e-08	3.475e-11)
%J1932+2220  (2.414e-08	1.129e-11).

\begin{table*}[ht]
% \tabletypesize{\tiny} no apparent effect
\begin{center}
%\tablewidth{0pt}
\caption{Twelve young (top) and four recycled (bottom) radio pulsars for which we see gamma-ray pulsations.
The ``Ref.'' column indicates the discovery paper.
Distances are taken from the ATNF pulsar catalog. %% ``DIST'' and ``DIST\_DM1'' variables.
The first is derived from the dispersion measure (DM) using the YMW16 model \citep{ymw16} except for J1932+2220,
which uses the kinematic distance of \cite{FrailWeisbergDistances}, and for J0636+5129, for which \citet{GBNCC-I} measured timing parallax. 
The second distance comes from the DM using NE2001 \citep{Cordes2002}.
The additional distance for J1731$-$4744 is from \citet{PM_B1727}.
The additional distance for J1946+3417 comes from the discussion of Figure \ref{Edot_sqEDOTD2}, see text.
The rotation ephemeris used in this work (last column) was obtained from radio timing with: 
N (Nan\c cay Radio Telescope) \cite{Cognard2011} ; 
P (Parkes Radio Telescope) \cite{ParkesFermiTiming} ; 
J (Jodrell Bank Observatory) \cite{Jodrell};
E (Effelsberg) \cite{J1946p3417_Barr} ;
G (Green Bank Telescope) \cite{GBT}.
The observatories in parentheses also provided timing observations. 
The MSPs with $\dot E$ uncertainties are Shklovskii-corrected (see text). 
The 6 pulsars with a \dag \,  are co-located with LAT 8-year sources.
\label{tbl-charPSR}}
%% GBT (Green Bank Telescope) (reference?).
\begin{scriptsize}
%\begin{footnotesize}
\begin{tabular}{lrrrrcrrrc}
\hline \hline
   PSR        &  Ref.\tablenotemark{a}   &   $l$  &  $b$  &  $P$   &  Distance    &$\dot E /10^{34}$&$S_{1400}$ &H-test & Timing  \\ %    &Years &Radio   
              &      &($^\circ$)&($^\circ$)&(ms)  &   (kpc)      & (erg s$^{-1}$)  & (mJy) & signif. & \\ %  &folded&discovery& \\
\hline
 J0729$-$1836 &  mlt+78 & 233.76 & -0.34 &  510.2 & 2.4 / 2.9    &   0.56    &  1.4  & 41   & P (J) \\ 
 J1731$-$4744\dag& lvw68& 342.56 & -7.67 &  829.8 & 5.5 / 2.8 (0.7)& 1.13    & 12.0  & 34   & P \\ 
 J1740+1000\dag  & mca00&  34.01 & 20.27 &  154.1 & 1.2 / 1.2    &  23.16    &  9.2  & 31   & J (N) \\ 
 J1757$-$2421 &  kom74  &   5.28 &  0.05 &  234.1 & 3.1 / 4.4    &   4.00    &  3.9  & 29   & P (J, N) \\ 
 J1816$-$0755 &  lfl+06 &  21.87 &  4.09 &  217.6 & 3.1 / 2.8    &   2.48    &  0.17 & 54   & J \\ 
 J1841$-$0524 &  hfs+04 &  27.02 & -0.33 &  445.7 & 4.1 / 5.3    &  10.42    &  0.2  & 27   & J (P) \\ 
 J1853$-$0004\dag&hfs+04&  33.09 & -0.47 &  101.4 & 5.3 / 7.1    &  21.09    &  0.87 & 43   & P (J, N) \\ 
 J1913+1011   &  mhl+02 &  44.48 & -0.17 &   35.9 & 4.6 / 4.8    & 287.14    &  0.5  & 32   & J \\ 
 J1925+1720   &  lbh+15 &  52.18 &  0.59 &   75.7 & 5.1 / 6.9    &  95.42    &  0.07 & 37   & J \\ 
 J1928+1746   &  cfl+06 &  52.93 &  0.11 &   68.7 & 4.3 / 5.8    & 160.39    &  0.28 & 31   & J (N) \\ 
 J1932+2220\dag& ht75b  &  57.36 &  1.55 &  144.5 & 10.9/ 7.5    &  75.38    &  1.2  & 44   & J \\ 
 J2208+4056\dag& slr+14 &  92.63 &-12.09 &  637.0 & 1.0 / 0.8    &  0.080    &  0.48 & 50   & J \\ 
   \\
 J0636+5129    & slr+14 & 163.91 & 18.64 &   2.86 & 0.2 / 0.5    &   $0.569 \pm 0.001$ &      & 48 & N \\ 
 J1125$-$6014\dag&fsk+04& 292.50 &  0.89 &   2.63 & 1.0 / 1.5    &   $0.45 \pm 0.3$     &  0.05 & 35 & P \\ 
 J1327$-$0755  & blr+13 & 318.38 & 53.85 &   2.68 & 25.0/ 1.7    &    0.36             &      & 25 & N (G) \\
 J1946+3417    & bck+13 &  69.29 &  4.71 &   3.17 & 7.0 / 5.2 (2.5)   &   $0.2 \pm 0.1$     & 0.29 & 29 & J  (E) \\
\hline
\end{tabular}
\tablenotetext{a}{bck+13: \citet{NHTRU-I}, blr+13: \citet{GBTdriftI}, cfl+06: \citet{PALFA-I}, fsk+04: \citet{PMPS-V}, hfs+04: \citet{Hobbs_pmbs}, ht75b: \citet{ht75b}, 
kom74: Komesaroff, M. M. 1974, unpublished, lbh+15: \citet{PALFA-IV}, lfl+06: \cite{PMPS-VI}, lvw68: \citet{EarlyMolonglo}, mca00: \citet{mca00}, 
mhl+02: \citet{Morris02_parkes}, mlt+78: \citet{Molonglo}, slr+14: \citet{GBNCC-I}.}
\end{scriptsize}
%\end{small}
\end{center}
\end{table*}

\section{A gamma-ray emission deathline} \label{DeathSection}
Striking in Figure \ref{H36_EDOT} is a minimum spindown power $\dot E$ for gamma-ray pulsars.
Most with $\dot E < 10^{34}$ erg s$^{-1}$ are MSPs.
The discovery of gamma-ray pulsations from PSR J2208+4056 establishes that while it may be rare for such low-power young pulsars to be seen with the LAT, 
it is not impossible. We discuss this pulsar further, below.

This new sample allows checks on attempts to understand which pulsars will or will not be seen in gamma rays.
\citet{DetectedNondetected} focused on the radio profiles of young pulsars with $\dot E > 10^{35}$  erg s$^{-1}$,
with the idea that the profiles result from the inclinations of the pulsar's rotation and magnetic axes relative to the line-of-sight
($\zeta$ and $\beta$, respectively).
We have seven such pulsars in Table \ref{tbl-charPSR}, five of which are in their sample.
In their Figure 2, radio pulse width $w_{10}$ versus $\dot E$, gamma-ray detected pulsars lie mostly below a diagonal 
$w_{10} = 48 \log \dot E - 1674$, for $\dot E$ in units of erg s$^{-1}$. Our seven straddle this line. 
Nevertheless, the general trend they pointed out -- that the third of high-$\dot E$ pulsars that remain undetected in gamma rays tend to be those with
broader radio pulses -- is borne out. 

For the MSPs, \citet{POND} used timing parallax distances, with well-determined uncertainties,
to confirm that once relative motion effects have been taken into account via the Shklovskii proper motion corrections,
the observed $\dot E \approx 10^{33}$ erg s$^{-1}$ MSP deathline value is robust.
The four new gamma-ray MSPs reported here are consistent with this.

Again for MSPs, \citet{gt+14} also look for observations that constrain the inclination angles, to then predict whether a gamma-ray
beam might sweep the Earth. 
During the MSP recycling process (i.e., during spin-up), the mass of the precursor of the white dwarf companion determines the orbital
radius, and thus the orbital period $P_B$. Using $P_B$ and $a_p \sin i$, they deduce $i$, the orbital inclination.
They use $i \approx \zeta$, and expect that gamma rays can be seen mainly for large $\zeta$, since emission models generally favor equatorial
gamma-ray beams. Three of our four MSPs have $P_B > 1$ day, the limit for their model's validity. For two, the $(P_B, a_p \sin i)$ values
lead to $\zeta$ values near $45^\circ$, and for the third $\zeta \approx 70^\circ$. 
These values fill in the $\zeta$ histogram of their Figure 5, tending to support their conclusions.

\subsection{PSR J2208+4056}
\label{J2208sect}
PSR J2208+4056 was originally called J2207+40 when discovered in the Green Bank North Celestial Cap Pulsar Survey \citep{GBNCC-I}.
An improved rotation ephemeris by \citet{GBNCC-III} revealed gamma pulsations directly.
Its location well off the Galactic plane, with no other LAT 8-year sources within $1^\circ .5$, ensures a low background level
and eases detection.

PSR J2208+4056 is the only gamma-ray detection out of 200 pulsars folded in the $10^{32} < \dot E < 10^{33}$ erg cm$^{-1}$ decade,
for a detection rate of $0.5^{+0.1}_{-0.4}$\%.
Scrutiny of its line-of-sight using the method of Figure 4 in \citet{SixWeak} suggests that its $\sim 0.9$ kpc DM distance is reliable.
Inspection of Figure \ref{Edot_sqEDOTD2} shows that $<30$ of the folded pulsars are `nearby', that is, $G_h$ is above the detectable minimum.

PSRs J2208+4056 and J1816$-$0755 are the fifth and sixth radio-interpulse pulsars seen in gamma rays.
Radio beams from both magnetic poles of the neutron star sweeping the Earth
implies that the pulsars are ``orthogonal rotators'', that is, that both the magnetic inclination $\alpha=\zeta-\beta$ and the angle $\zeta$ from
the line-of-sight to the rotation axes are near $90^\circ$ \citep{interpulses}. We thus observe the neutron star's equator,
where the gamma-ray beam is most intense and $f_\Omega < 1$ \citep{Watters09}, in addition to seeing both magnetic poles sweep by.
Models further predict that equatorial emission is enhanced as pulsars age.  With characteristic ages of $\tau_c = 1.9$ and $0.54$ Myr, these two
pulsars are among the older ones seen with the LAT.  These pulsars' geometry thus favor detection.

\citet{PolnEnergeticPSRs} show that radio polarization in young pulsars is markedly stronger for $\dot E \ga 10^{34}$ erg s$^{-1}$.
\citet{PolnPSRs2018} confirmed the trends using 600 pulsars.
It seems an unlikely coincidence that both radio polarization and gamma-ray emission require the same minimum spin parameters 
for independent reasons. 
We observed PSR J2208+4056 at 1400 MHz with the Nan\c cay Radio Telescope and obtained a preliminary estimate of 50\% linear polarization.
In Figure 8 of \citet{PolnEnergeticPSRs}, this places the pulsar among the most polarized for that $\dot E$ range.
Ideally, an interpulse allows good fits of polarization position angle versus phase, with well-defined $\alpha, \zeta$ values in consequence.
Unfortunately, the 1400 MHz flux densities for PSRs J1816$-$0755 and J2208+4056 are 170 and $<800\,\mu$Jy, respectively, 
and the interpulses are $<40$\% as intense as the main pulses.
Position angle fits to both pulses for either pulsar will be a challenge. Observations at lower frequencies are of interest.

In summary, PSR J2208+4056 establishes that the minimum spindown power for gamma-ray emission is lower than previously observed.
Since $L_\gamma$ for low-power pulsars is low, detection requires favorable geometry and sky location,
partially explaining the low ($\sim 0.5$\%) detection rate. 
\subsection{Distance Biases}
With our large sample of pulsars we can explore whether the apparent deathline is an artifact of a distance-related selection bias. 
Figure \ref{Edot_sqEDOTD2} shows $G_h$ % = \sqrt{\dot E} /(4\pi d^2)$
versus the spindown power $\dot E$, for our twelve hundred pulsars. 
$G_h$ is our best predictor of gamma-ray detectability since, as shown in 2PC Figure 9,
gamma-ray luminosity scales roughly as $L_\gamma \propto \sqrt{\dot{E}}$.
\citet{Arons96} predicted the correlation, arguing that the open field-line voltage is $V \simeq 3.18 \times 10^{-3} \sqrt{\dot E }$ volts,
and that above some threshold voltage, the size of the gamma-ray emitting electron-positron cascades
increases linearly with $V$.

Except for some outliers, the minimum $G_h$ for known gamma-ray pulsars in Figure \ref{Edot_sqEDOTD2} is  
$G_h^\mathrm{min} = 10^{15}$ (erg s$^{-1})^{1/2}$ kpc$^{-2} = 3.3\times 10^{-12}$ ${\rm erg} \, {\rm cm}^{-2} \, {\rm s}^{-1}$.
The figure seems to show that if pulsars with  $\dot E < 10^{33}\, {\rm erg \,s}^{-1}$ emit gamma rays, 
then we would have detected some of those with high $G_h$. 
However, Figure 3 of \citet{LaffonNewPSRs} shows the fraction of folded pulsars detected in gamma rays, as a function of $\dot E$.
Updating the curves with the recent discoveries preserves the overall trends.
For young pulsars, the fraction is $\sim 60$\% for $\dot E > 10^{35}\, {\rm erg \,s}^{-1}$, dropping to $\sim 5$\% for the decade 
$ 10^{33} < \dot E < 10^{34}\, {\rm erg \,s}^{-1}$. 
With the detection of PSR J2208+4056, we now have $0.5$\% detected in the decade below that.
For MSPs, 40\% of the pulsars in the low decade are seen, rising to $>80$\% for high $\dot E$.
Hence, the detection efficiency at low $\dot E$ is small but non-zero.

Below $\dot E < {10^{33}\, {\rm erg \,s}^{-1}}$ Figure \ref{Edot_sqEDOTD2} shows 48 (46) pulsars with $G_h > G_h^\mathrm{min}$ that we have (not) phase-folded.
Over 70\%, folded or not, are off the plane, $|b| > 5^\circ$, where the diffuse background is low.
The low efficiency, applied to the small sample, predicts effectively no detections.
Low-power pulsars thus may emit gamma rays, but there are not enough within distances accessible by the {\em Fermi} LAT for us to know.
It will be interesting to add a population of old pulsars to a population synthesis to see whether
the low flux per pulsar combined with the large population of low power pulsars adds a noticeable contribution
to the diffuse Galactic emission.
%
%Plotting the fractions instead as a function of $G_h$ shows a linear increase (instead 
%Golly, the min of $\sqrt{\dot E} / 4\pi d^2$ was $10^{16}$ in 2PC but here is below $10^{15}$.
%
\subsection{Spectra}
We evoked, above, cascades of electrons and positrons accelerated by high voltages in the regions
where the magnetic field lines open as they cross the light cylinder.
The maximum gamma-ray energies from curvature radiation are higher when the field-line radii of curvature are smaller. 
These radii decrease with rotation period, because the light cylinder is smaller,
and when the magnetic inclination $\alpha$ is large. A stronger magnetic field induces higher $V$ for a given period,
and thus bigger cascades and, as long as the resulting plasma has not shorted out the electric field, higher $L_\gamma$. 
In this picture, the gamma-ray spectral shape -- intensity, hardness (spectral index $\Gamma$), and cutoff energy $E_\mathrm{cut}$ -- depend on the spindown parameters. 
Modelers aim to refine and quantify this picture.

In particular, Figure 2 of \citet{Kalapotharakos2017} builds upon an empirical result from 2PC,
that $\log E_\mathrm{cut}$ is a parabolic function of $\log \dot E$. 
Figure 7 in 2PC shows $\Gamma$ increasing linearly with $\dot E$, and we discussed $L_\gamma \propto \sqrt{\dot E}$ above.
\citet{Kalapotharakos2017} suggest that combining the three correlations causes the energy of greatest flux to scale with $\dot E$. 
%gives a family of spectral shapes as shown in the left-hand frame of Figure \ref{FluxVsCutoff}.
The deathline would be due not to the emission mechanism abruptly shutting off below some minimum $\dot E$,
but rather to a combination of reduced intensity and, importantly, 
emission shifting to the MeV range where the LAT is less sensitive. 
As a counterpoint, we mention that the MeV pulsars in \citet{KuiperHermsen2015} mostly have very high $\dot E$.
The upcoming \textit{Fermi} Third Pulsar Catalog will double the number of measured spectra as compared to 2PC,
allowing better constraints on how spectral shapes depend on spin parameters.

%\begin{figure}[ht!]
%\centering
%\includegraphics[width=0.9\textwidth, angle=0]{FluxVsCutoff_steeper.eps}
%\caption{A toy model of photon flux versus spindown power. 
%Left: Typical Spectral Energy Distributions (SEDs) 
%$E_\gamma^2 \frac{{\rm d} N}{{\rm d} E_\gamma} $
%for different $\dot E$ values,
%assuming that $\Gamma$, $E_{\rm cut}$, and luminosity evolve with $\dot E$  as suggested by the 2PC results.
%Right: Photon fluxes above 50 and 100 MeV obtained by integrating $\frac{{\rm d} N}{{\rm d} E_\gamma} $
%for each  $\dot E$.
%A shift in the SED peak to lower energies with decreasing $\dot E$ on its own does not explain the sharp observed
% $\dot E$  deathline, a break in  $L(\dot E)$ is also necessary.
%This plot may not make it to the final paper, or may evolve. 
%\label{FluxVsCutoff}}
%\end{figure}

\section{Summary} \label{ConclusionsSection}
In its $11^{th}$ year on orbit, {\em Fermi} LAT continues to discover about 24 gamma-ray pulsars per year.
Two discovery channels are deep radio searches and blind gamma-ray periodicity searches at the
positions of bright pulsar-like unidentified gamma-ray sources. This paper concerns a third channel,
phase-folding gamma rays using a rotation ephemeris obtained from sustained observations of previously known pulsars.
This channel is the most sensitive to gamma-faint pulsars.

Gamma-ray pulsations for 16 pulsars were presented, 12 young and 4 recycled, made possible because 
i) several hundred updated (1269 in total) ephemerides were applied to a $9.6$ year gamma-ray data set ; 
ii) \citet{SearchPulsation}'s gamma-ray weighting technique can be applied regardless of whether or not the pulsar is seen as a phase-integrated point source ; 
iii) we demonstrated that the detection threshold can be lowered to $4.1\sigma$ without incurring false positive detections. 
The new gamma-ray pulsars include the faintest yet seen, giving fewer than ten photons per hundred days. 
We described some of the consequences of the resulting large Poisson fluctuations.
PSR J2208+4056 has $\dot E = 8\times 10^{32}\, {\rm erg \,s}^{-1}$, three times lower than what had been the lowest spindown powers
for known gamma-ray pulsars these last several years.

We discussed selection biases associated with the new discoveries. Most of the new pulsars have very narrow peaks,
improving sensitivity, and lie at low Galactic latitude, where the background is intense. 
The Third Gamma-ray Pulsar Catalog (``3PC'', in preparation) will contain more than twice as many gamma-ray pulsars as 2PC.
The unambiguous identification and localisation of gamma-ray pulsars helps improve 4FGL, the upcoming $4^{th}$ {\em Fermi} LAT catalog of sources using 8 years of data.
Both catalogs should go to press in the coming year.

Emission modelers should bear in mind that since faint pulsars are easier to detect if the pulses are narrow,
tuning models to match the observed sample would be a mistake. 
If, for example, pulses broaden as $\dot E$ decreases, those pulsars would be absent from the current observed sample.

We discussed whether the rarity of detected gamma-ray pulsars with spindown power $\dot E < 10^{33}\, {\rm erg \,s}^{-1}$
is a feature of the gamma-ray emission mechanism (a ``deathline''), or is due to observational selection effects.
There does seem to be a deathline, with luminosity dropping below the $L_\gamma \propto \sqrt{\dot E}$ trend at low $\dot E$.
But faint emission at low $\dot E$ for rare pulsars may be undetected at present because 
the distances implied by the large volumes required to obtain a pulsar sample large enough to provide a few detections
impose values of $G_h = \sqrt{\dot E} /(4\pi d^2)$ below the LAT's sensitivity.
Sensitivity is enhanced for narrow pulses and for high Galactic latitudes, but requiring those conditions
decreases the sample size. We also mentioned model predictions that peak emission would shift from the GeV to the MeV
energy range as $\dot E$ decreases, making LAT detections more difficult. 

Along with the simple scheme applied here, 
\citet{SearchPulsation} presents a pulsar weighting method that is more sensitive but also more difficult to use.
Future work will be to re-fold the thousand pulsars, as well as other pulsars for which ephemerides may become available, with the refined method.
Preliminary indications are that we will find still more faint gamma-ray pulsars.

%Finding the contribution of the unresolved population to diffuse emission has turned out to be important for the battle of the bulge,
%and also for M31 maybe, and for globulars.
%(cite pierbattista and the atlases)

{\bf Acknowledgements}

The Nan\c cay Radio Observatory is operated by the Paris Observatory, associated with the French Centre National de la Recherche Scientifique (CNRS). 

The Parkes radio telescope is part of the Australia Telescope which is funded by the Commonwealth Government for operation as a National Facility managed by CSIRO. 

The Lovell Telescope is owned and operated by the University of Manchester as part of the Jodrell Bank Centre for Astrophysics with 
support from the Science and Technology Facilities Council of the United Kingdom.

The Robert C. Byrd Green Bank Telescope (GBT) is operated by the National Radio Astronomy Observatory, a facility of the National Science Foundation operated under cooperative agreement by Associated Universities, Inc.

The \textit{Fermi}  LAT Collaboration acknowledges generous ongoing support from a number of agencies and institutes that have supported both 
the development and the operation of the LAT as well as scientific data analysis. 
These include the National Aeronautics and Space Administration and the Department of Energy in the United States, 
the Commissariat \`a l'Energie Atomique and the Centre National de la Recherche Scientifique / Institut National de Physique Nucl\'eaire et de Physique des Particules 
in France, the Agenzia Spaziale Italiana and the Istituto Nazionale di Fisica Nucleare in Italy, 
the Ministry of Education, Culture, Sports, Science and Technology (MEXT), High Energy Accelerator Research Organization (KEK) 
and Japan Aerospace Exploration Agency (JAXA) in Japan, and the K.~A.~Wallenberg Foundation, 
the Swedish Research Council and the Swedish National Space Board in Sweden.

Additional support for science analysis during the operations phase is gratefully acknowledged from the 
Istituto Nazionale di Astrofisica in Italy and the Centre National d'\'Etudes Spatiales in France. 
This work performed in part under DOE Contract DE- AC02-76SF00515.
Work at NRL is supported by NASA.

\bibliographystyle{aasjournal}

\bibliography{2ndPulsarCatalog}

\end{document}